\documentclass[aps,prd,amsmath,twocolumn,superscriptaddress,preprintnumbers,amssymb,showpacs,floatfix,nofootinbib,longbibliography]{revtex4-1}
\pdfoutput=1

\usepackage[T1]{fontenc}
\usepackage{bm}
\usepackage{graphicx}
\usepackage{float}
\usepackage{bm}
\usepackage{hyperref}
\usepackage{aas_macros}
\usepackage{multirow}
\usepackage{array} 
\usepackage{color,xcolor}
\usepackage{caption}
\usepackage{subcaption}
\usepackage{amsmath,amssymb}

\newcommand{\beq}{\begin{equation}} \newcommand{\eeq}{\end{equation}}
\newcommand{\bea}{\begin{eqnarray}} \newcommand{\eea}{\end{eqnarray}}

\definecolor{colorRKL}{rgb}{.6,.1,.5}

\hypersetup{
  colorlinks   = true, 
  urlcolor     = magenta, 
  linkcolor    = magenta, 
  citecolor   = magenta 
}

\begin{document}

\preprint{SLAC-PUB-17679}
\preprint{IPPP/22/35}

\title{
Tidal Love Numbers of Novel and Admixed Celestial Objects
}

\author{Michael Collier}
\thanks{\href{mailto:michael.collier@durham.ac.uk}{michael.collier@durham.ac.uk}; \href{http://orcid.org/0000-0003-3359-3706}{0000-0003-3359-3706}
}
\affiliation{Department of Physics, Durham University, Durham DH1 3LE, U.K.}

\author{Djuna Croon}
\thanks{\href{mailto:djuna.l.croon@durham.ac.uk}{djuna.l.croon@durham.ac.uk}; \href{http://orcid.org/0000-0003-3359-3706}{0000-0003-3359-3706}}
\affiliation{Department of Physics, Durham University, Durham DH1 3LE, U.K.}
\affiliation{Institute for Particle Physics Phenomenology, Department of Physics, Durham University, Durham DH1 3LE, U.K.}

\author{Rebecca K. Leane}
\thanks{\href{mailto:rleane@slac.stanford.edu}{rleane@slac.stanford.edu}; \href{http://orcid.org/0000-0002-1287-8780}{0000-0002-1287-8780}}
\affiliation{Particle Theory Group, SLAC National Accelerator Laboratory, Stanford University, Stanford, CA 94035, USA}
\affiliation{Kavli Institute for Particle Astrophysics and Cosmology, Stanford University, Stanford, CA 94035, USA}

\begin{abstract}
A sub-fraction of dark matter or new particles trapped inside celestial objects can significantly alter their macroscopic properties. We investigate the new physics imprint on celestial objects by using a generic framework to solve the Tolman-Oppenheimer-Volkoff (TOV) equations for up to two fluids. We test the impact of populations of new particles on celestial objects, including the sensitivity to self-interaction sizes, new particle mass, and net population mass. Applying our setup to neutron stars and boson stars, we find rich phenomenology for a range of these parameters, including the creation of extended atmospheres. These atmospheres are detectable by their impact on the tidal love number, which can be measured at upcoming gravitational wave experiments such as Advanced LIGO, the Einstein Telescope, and LISA. We release our calculation framework as a publicly available code at \href{https://zenodo.org/record/6584578\#.YpNkwlTMJhE}{this URL}, allowing the TOV equations to be generically solved for arbitrary new physics models in novel and admixed celestial objects.
\end{abstract}

\maketitle

\section{Introduction}

Celestial objects are excellent new physics detectors. Their deep gravitational wells offer an opportunity to capture dark matter (DM) or other new particles, if the new particles lose enough energy through scattering with the Standard Model (SM) celestial matter. Such captured populations lead to an array of exciting signatures. If the captured particles annihilate and the products are absorbed, the celestial object can have an increased temperature~\cite{Goldman:1989nd,
Bertone:2007ae,
Mack:2007xj,
deLavallaz:2010wp,
Kouvaris:2010vv,
McCullough:2010ai,
Baryakhtar:2017dbj,
Raj:2017wrv,
Bell:2018pkk,
Chen:2018ohx,
Dasgupta:2019juq,
Hamaguchi:2019oev,
Camargo:2019wou,
Bell:2019pyc,
Acevedo:2019agu,
Joglekar:2019vzy,
Joglekar:2020liw,
Leane:2020wob,
Bell:2020jou,
Dasgupta:2020dik,
Garani:2020wge,
Bramante:2019fhi,Bell:2021fye,Freese:2008hb, Taoso:2008kw, Ilie:2020iup, Ilie:2020nzp}. If the products escape, SM particles such as gamma-rays, electrons, and neutrinos can be detected directly~\cite{Leane:2017vag, HAWC:2018szf, Nisa:2019mpb,Bell:2011sn,Super-Kamiokande:2015xms,IceCube:2016dgk,ANTARES:2016xuh,Leane:2021ihh,Leane:2021tjj,Bose:2021yhz}. If the new particles do not annihilate away, a large population can remain undepleted inside the celestial object and have dramatic consequences. One example is that a black hole may form at their heart due to overdensities, and they may consequently implode~\cite{PhysRevD.40.3221,Kouvaris:2007ay,Kouvaris_2010,Kouvaris:2011fi, deLavallaz:2010wp,McDermott:2011jp, Guver:2012ba,Kouvaris:2012dz,Bramante:2013hn,Bell:2013xk,Bramante:2013nma, Bramante:2014zca,Bramante:2015dfa,fan2012constraining}.

The first detection of gravitational waves in 2017 by LIGO/VIRGO has presented an opportunity to study celestial objects in a new band of the multi-wavelength sky. This opens up exciting new prospects to use this probe to search for new particle interactions. One testable scenario is that new long-range interactions between SM particles or DM in binary stellar systems can affect their inspiral, causing waveform corrections~\cite{Ellis:2017jgp,Croon:2017zcu}. Alternatively, if sufficiently large amounts of new particles are trapped, this may affect the macroscopic properties of the celestial body, such as its mass and radius, due a softened equation of state (EOS). This leads to two key physical observables. First, a reduction in the stellar mass, which can be compared with the heaviest known objects, to set a constraint~\cite{Li:2012ii, Leung:2011zz, Xiang:2013xwa,Tolos:2015qra,Panotopoulos:2017idn,Gresham:2018rqo,Karkevandi:2021ygv}. Second, a tidal deformability, quantified by a "Love number", which is detectable in gravitational waves through a phase shift~\cite{Nelson:2018xtr,Dengler:2021qcq}.

Observing any macroscopic change in the celestial object's properties, through a gravitational wave signal or otherwise, generally requires a large abundance of trapped DM. This DM can be implanted into the object at its birth, for example neutron stars may retain new populations produced in their origin supernova~\cite{Ellis:2018bkr,Nelson:2018xtr}. It is also possible that new particle compact objects or clumps can accrete baryonic matter, leading to a large additional core in celestial objects~\cite{Ciarcelluti:2010ji}. Capture also may occur from a dark companion star~\cite{Ciarcelluti:2010ji}. 

The goal of this work is to study the sensitivity of the properties of celestial objects to new physics parameters, including DM or new particle self-interactions. We will remain agnostic to the precise new particle production or trapping mechanism, and instead focus on the new phenomenology and distribution of these particles around objects, and prospects for detecting these features. To do this, we solve the Tolman-Oppenheimer-Volkoff (TOV) equations for up to two fluids. We investigate resultant phenomenology, including dark atmospheres, which can have large extent outside of the celestial object's radius. These dark atmospheres can impact binary star systems, as the internal degrees of freedom of the bodies can appreciably influence their inspiral. This allows for a direct probe of the celestial body EOS, which can be measured at upcoming gravitational wave experiments such as the Einstein Telescope~\cite{Gupta:2022qgg}, Advanced LIGO/Virgo/KAGRA or LISA \cite{LISA:2022kgy}. As an important component of this work, we release our calculation framework as a publicly available code, allowing the TOV equations to be generically solved for arbitrary new physics models and a range of (admixed) celestial objects.

This paper is organized as follows. We begin by detailing equations of state in Section~\ref{sec:eos}, and demonstrate the new physics profiles and stability for boson stars and admixed neutron stars in Section~\ref{sec:profiles}. We then study the sensitivity of upcoming gravitational wave experiments to admixed neutron star and boson star observables such as the tidal love number in Section~\ref{sec:love}. We briefly review and discuss some example applications of these results in Section~\ref{sec:appl}, and conclude and summarize our results in Section~\ref{sec:conclusion}.

\section{Equations of State and Macroscopic Parameters}
\label{sec:eos}

We detail our equations of state (EOS) and framework to solve the TOV equations for up to two fluids, which we will later apply to a new particle population inside either neutron stars, or a pure boson star.

\subsection{The TOV Equations}

Macroscopic parameters such as the mass and radius of a neutron star can be found by solving the TOV equations, of the form (in natural units),
\begin{equation}
    P' =-\frac{G_N (m  \rho ) \left(\frac{ P}{\rho }+1\right) \left(\frac{4 \pi  r^3  P}{m }+1\right)}{r^2 \left(1-\frac{G_N (2 m )}{r}\right)},
\end{equation}
where $P$ is the pressure, $\rho$ is the density, $r$ is the radius, $m$ is the mass, and the prime denotes a derivative with respect to the given parameter.  We will also allow for an admixture of two different forms of matter which interact gravitationally. In this case, the set of coupled TOV equations is given by~\cite{Oppenheimer:1939ne, Tolman:1934za}
\begin{equation}
    \begin{split}
         P_{\rm 1}' &=-\frac{G_N (m  \rho_{\rm 1} ) \left(\frac{ P_{\rm 1}}{\rho_{\rm 1} }+1\right) \left(\frac{4 \pi  r^3 ( P_{\rm 2}+ P_{\rm 1})}{m }+1\right)}{r^2 \left(1-\frac{G_N (2 m )}{r}\right)},
         \\
         P_{\rm 2}'  &= -\frac{G_N (m  \rho_{\rm 2} ) \left(\frac{ P_{\rm 2}}{\rho_{\rm 2} }+1\right) \left(\frac{4 \pi  r^3 ( P_{\rm 2}+ P_{\rm 1})}{m }+1\right)}{r^2 \left(1-\frac{G_N (2 m )}{r}\right)},
         \\ 
         m'  &= 4 \pi  r^2 (\rho_{\rm 2} +\rho_{\rm 1} ),
    \end{split}
\end{equation}
where indices 1 and 2 refer to the two different fluids. These two fluids for example could consist of firstly a nuclear matter component, and the second component a fermionic field belonging to a hidden sector. 
We now describe the various EOS we will use in this work. 

\subsection{Equations of State}
\subsubsection{Nuclear EOS}
Various models have been proposed to describe nuclear matter properties in dense environments, in particular at super-nuclear densities $n > n_0 \sim 0.16 \,{\rm fm}^{-3}$. 
Such descriptions are either based on a Hamiltonian (the potential models), or on a Lagrangian (the field-theoretical models).
Here we will adopt the Brussels-Montreal functional BSk22 \cite{Pearson:2018tkr} (see Ref.~\cite{Goriely:2013xba} for BSk19-21), a nuclear EOS with parameters determined primarily by fitting to the measured masses of atomic nuclei having $Z$, $N \geq 8$ from the 2012 Atomic Mass Evaluation \cite{audi2012ame2012}.
BSk22 is relatively stiff, and able to produce heavy neutron stars as consistent with observations. Note that this EOS is simply a benchmark serving the purpose of demonstrating the impact of new physics, it has own sets of constraints, and other EOS may be favored further in future (see e.g. Ref.~\cite{CHAMEL_2013}). 

\begin{figure}[t!]
    \includegraphics[scale = 0.5]{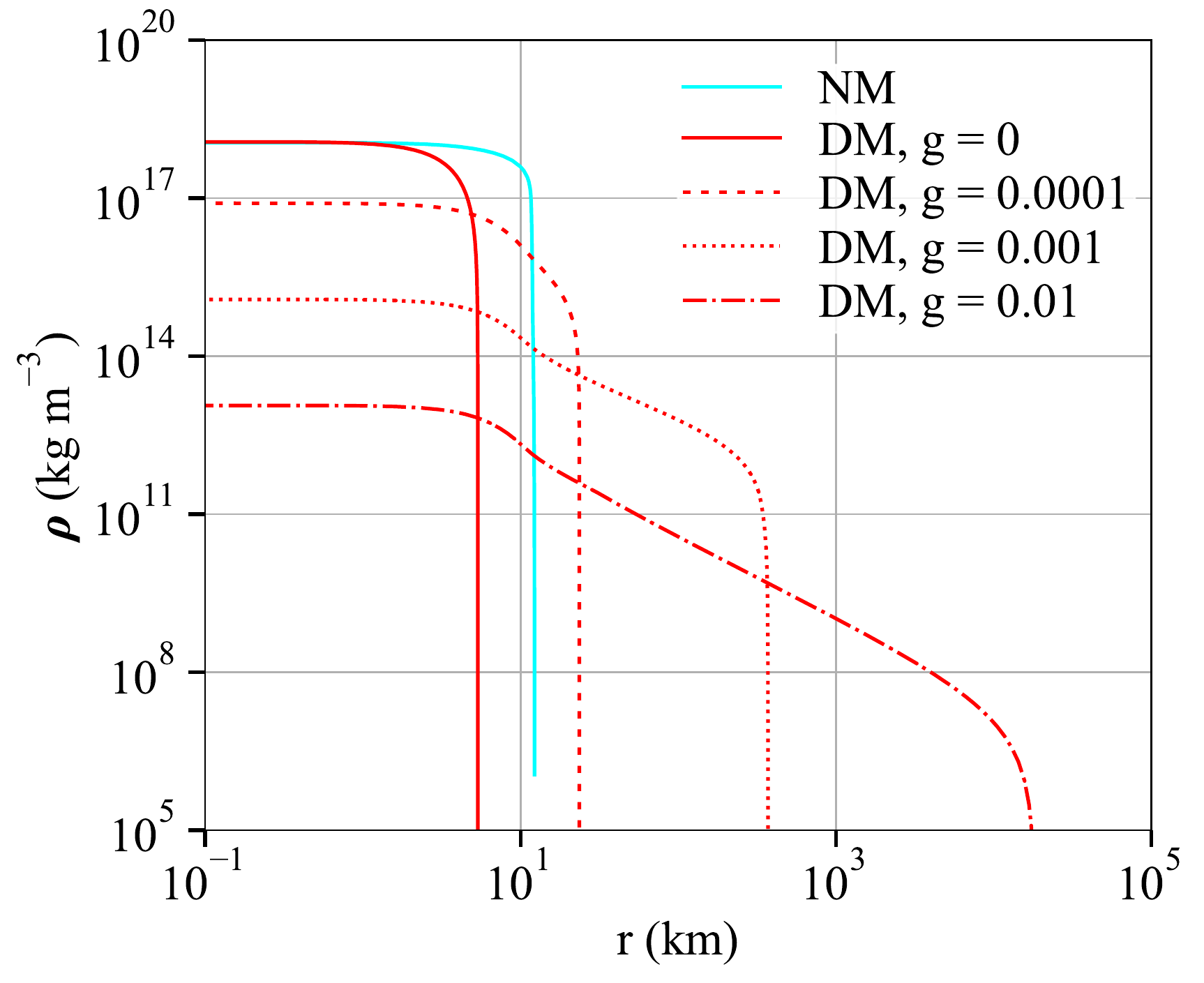}
    \caption{Density profile $\rho$ as a function of radius $r$ for a neutron star composed of nuclear matter (NM) with a DM or new particle with mass 1 GeV which makes up $5\%$ of the neutron star mass, a self-interaction mediator with mass 1 keV, a central density of $\rho_{c} = 2.3 \times 10^{18} \mathrm{\, kg \, m^{-3}},$ and a range of self couplings $g$.}
    \label{fig:profile}
\end{figure}

\subsubsection{Fermions with Yukawa Interactions}
To study the effects of new particles, we will consider fermionic matter with an equation of state given by \cite{Kouvaris:2015rea, Mukhopadhyay:2016dsg}
\begin{equation}
\label{eq:ferm}
    \begin{split}
        \rho =& \frac{m_\chi^4}{8 \pi^2} \left( x \sqrt{1+x^2} (1 + 2 x^2) - \ln\left(x + \sqrt{1+x^2} \right)
        \right) \\
        &+ \frac{g^2 x^6 m_\chi^6}{2 (3 \pi^2)^2 m_\phi^2}
        \\
        P =& \frac{m_\chi^4}{8 \pi^2} \left( x \sqrt{1+x^2} \left( \frac{2 }{3} x^2 -1\right) + \ln\left(x + \sqrt{1+x^2} \right)
        \right)\\
        &+ \frac{g^2 x^6 m_\chi^6}{2 (3 \pi^2)^2 m_\phi^2}
    \end{split}
\end{equation}
where we work in natural units $\hbar = c = 1$, and where $ x = p/m_\chi$. Here $g$ is the coupling between the fermion and the mediator $\phi$, $m_\phi$ is the mediator mass, and $m_\chi$ is the particle mass. The last terms in Eq.~(\ref{eq:ferm}) correspond to the contribution from self-interactions. Note that this EOS is consistent for the repulsive self-interactions as we consider, however can be inconsistent for relativistic fermions in the case of attractive self-interactions, mediated by scalars~\cite{Walecka:1974qa,Gresham:2018rqo}. Furthermore, as we focus on repsulive interactions, we do not expect a BCS phase or BEC to form, but in the case of attractive interactions these would lead to a different EOS and interesting complementary effects~\cite{Schmitt:2014eka}.

\subsubsection{Bosons with Repulsive Self-Interactions}

The Bose-Einstein Condensate EOS is given by \cite{Li:2012qf}
\begin{equation}
    \label{eq:boson star EOS}
    P=\frac{\sqrt{\pi\sigma}}{m^3}\,\rho^2,
\end{equation}
where $P$ is the pressure, $\rho$ is the density, $m$ is the new particle mass and $\sigma$ its repulsive self-interaction cross section. We will use this EOS to describe hypothetical boson stars, which are effectively astrophysical Bose-Einstein condensates.

\begin{figure}[t!]
    \centering
    \includegraphics[scale = 0.55]{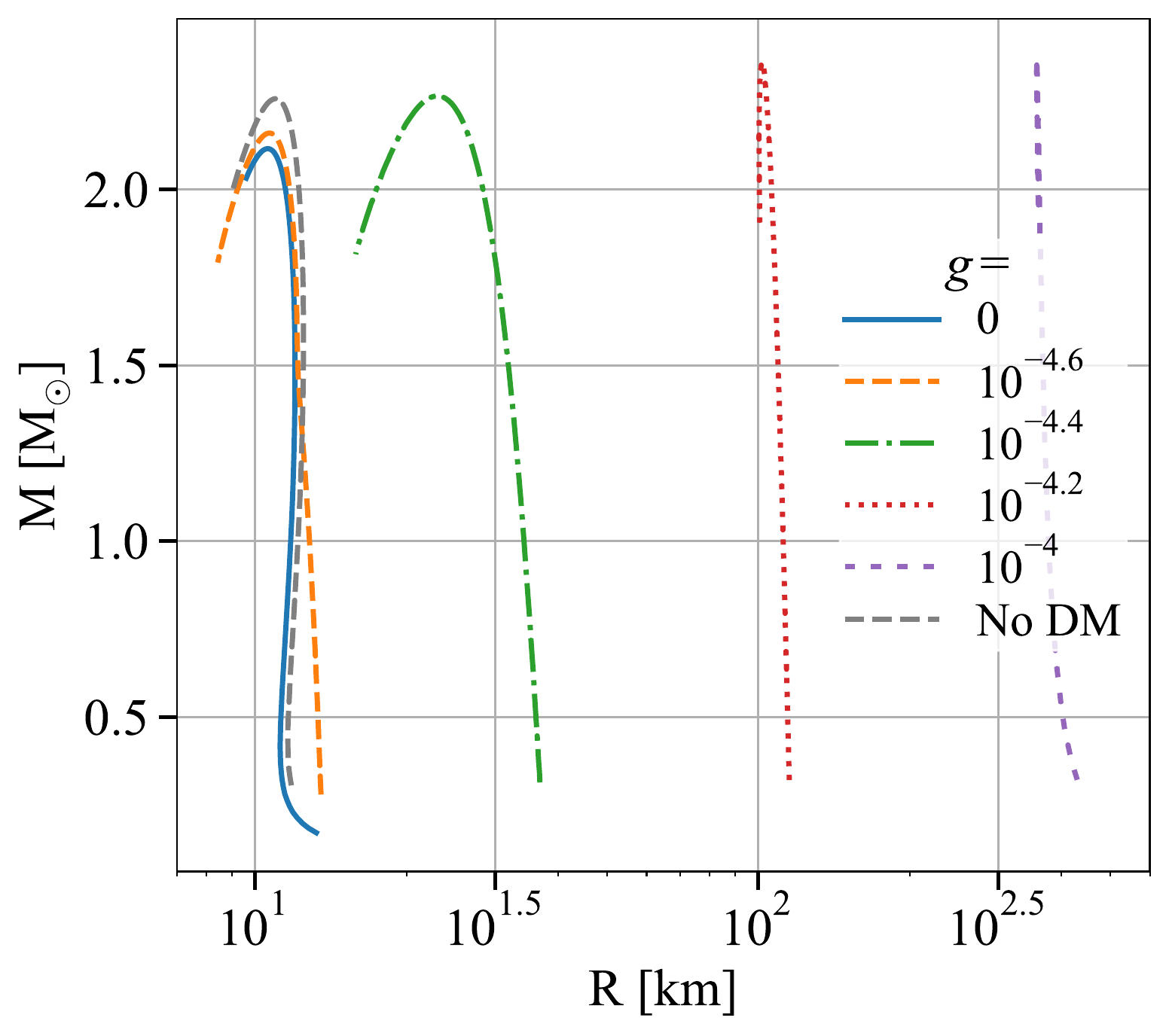}
    \label{fig:stability curves varied coupling R99}
  \caption{Mass-radius stability curves for neutron stars which are $5\%$ DM by mass, containing DM with varying values of the gauge coupling $g$. The DM has $m_\chi = 1 \, \mathrm{GeV}$ and $m_\phi = 1 \, \mathrm{keV}$.}
    \label{fig:stability_curves_varied_coupling}
\end{figure}

\begin{figure}[t!]
    \centering
    \includegraphics[scale = 0.55]{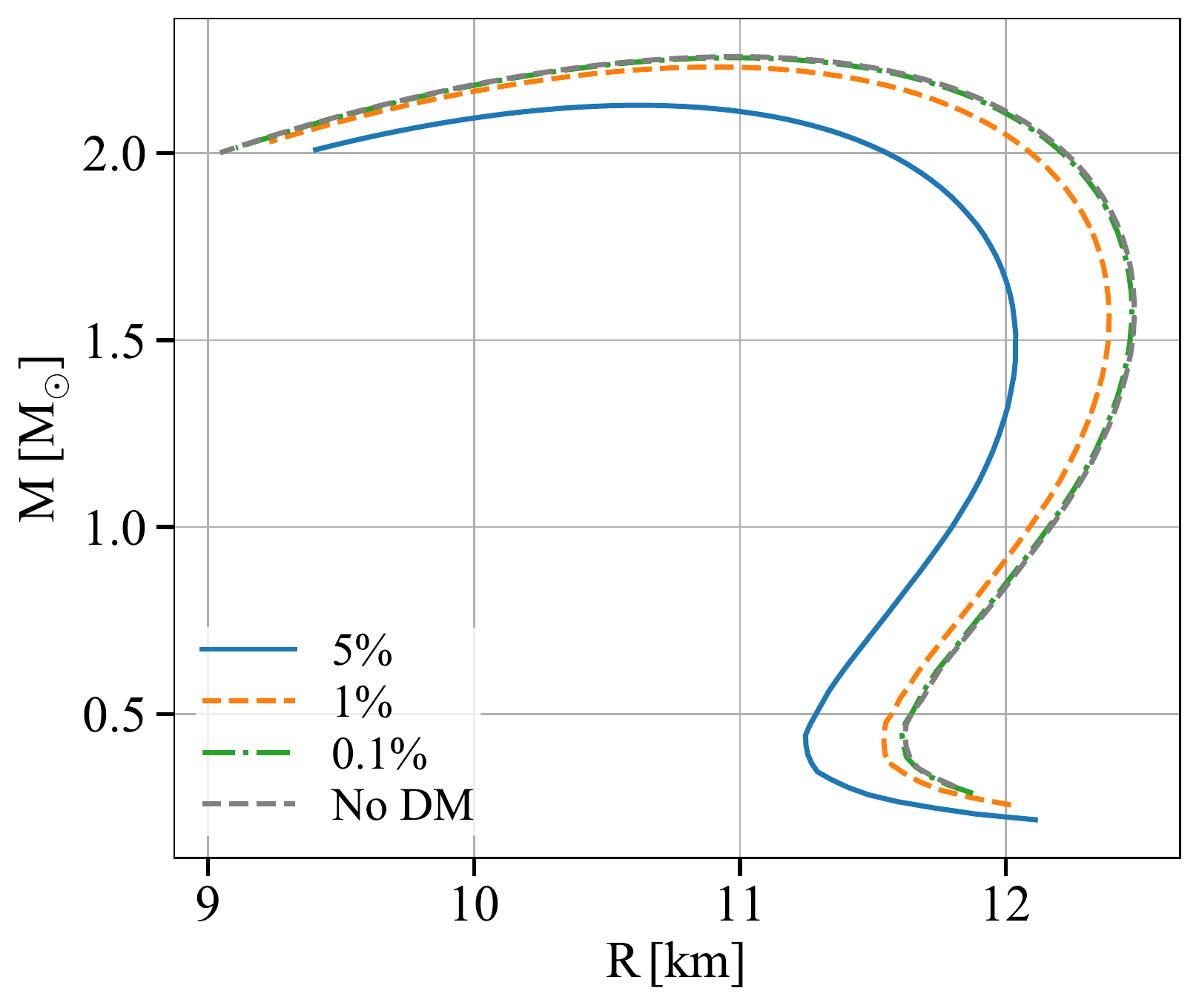}
    \caption{Mass - radius stability curve for a DM admixed neutron star with $g = 10^{-5}$  coupling and particle mass $m_\chi = 1 \, \mathrm{GeV}$. Curves are plotted at several different DM mass fractions as labeled.}
    \label{fig:stability_curves_varied_masspop}
\end{figure}

\section{Celestial Object Profiles and Stability}
\label{sec:profiles}

The presence of additional degrees of freedom in a celestial object can alter its physical properties. Remaining agnostic to the new physics production mechanism, we now examine the distributions of new physics or dark matter within two benchmark examples: an admixed neutron star, and a hypothetical pure boson star.

Figure~\ref{fig:profile} shows an example of a neutron star profile with a mass subfraction of new fermionic matter, assuming the BSk22 nuclear EOS and fermionic EOS described above. Here we show as a benchmark example neutron star containing a dark matter or new particle with mass 1 GeV which makes up $5\%$ of the neutron star by mass, and has self-interactions via a light mediating particle, taken as an example benchmark of mass 1 keV. We see that for increasing repulsive self-interaction, the dark matter has an increasingly puffy configuration, extending well outside the sphere of nuclear matter. On the other hand, no or little self-interaction leaves the dark matter settled in the core of the neutron star. These features are consistent with those found in earlier works, see e.g. Refs.~\cite{Nelson:2018xtr,Miao:2022rqj}.

These new features have important implications for the stability of the neutron star, as its macroscopic features such as mass and radius are clearly altered. Generally, these additional degrees of freedom will soften the EOS, leading to lighter mass neutron stars than would otherwise be expected. Therefore, such changes can be compared with the observation of the heaviest neutron star, the pulsar PSR
J0740+6620 at $2.14^{+0.10}_{-0.09}$~M$_\odot$~\cite{Cromartie:2019kug}, and potential constraints can be set on the abundance and properties of new physics in these objects.

Figure~\ref{fig:stability_curves_varied_coupling} and shows a range of expected mass-radius stability curves, which we obtain by solving the TOV equations detailed above. Here we take the same new physics benchmark parameters as Fig.~\ref{fig:profile}, although include even smaller self-couplings to demonstrate the expected stability behaviour. Consistent with Fig.~\ref{fig:profile} we see that increasing the repulsive self-interaction increases the total size of the neutron star.
Fig.~\ref{fig:stability_curves_varied_masspop} shows the scenario where instead the size of the new population is varied; increasing the population size decreases the maximum radius of the neutron star. For both cases, we show that a range of these self-interactions are compatible with the heaviest known neutron star.

\begin{figure}[t!]
    \hspace*{-0.5cm}
    \centering
    \includegraphics[scale=0.55]{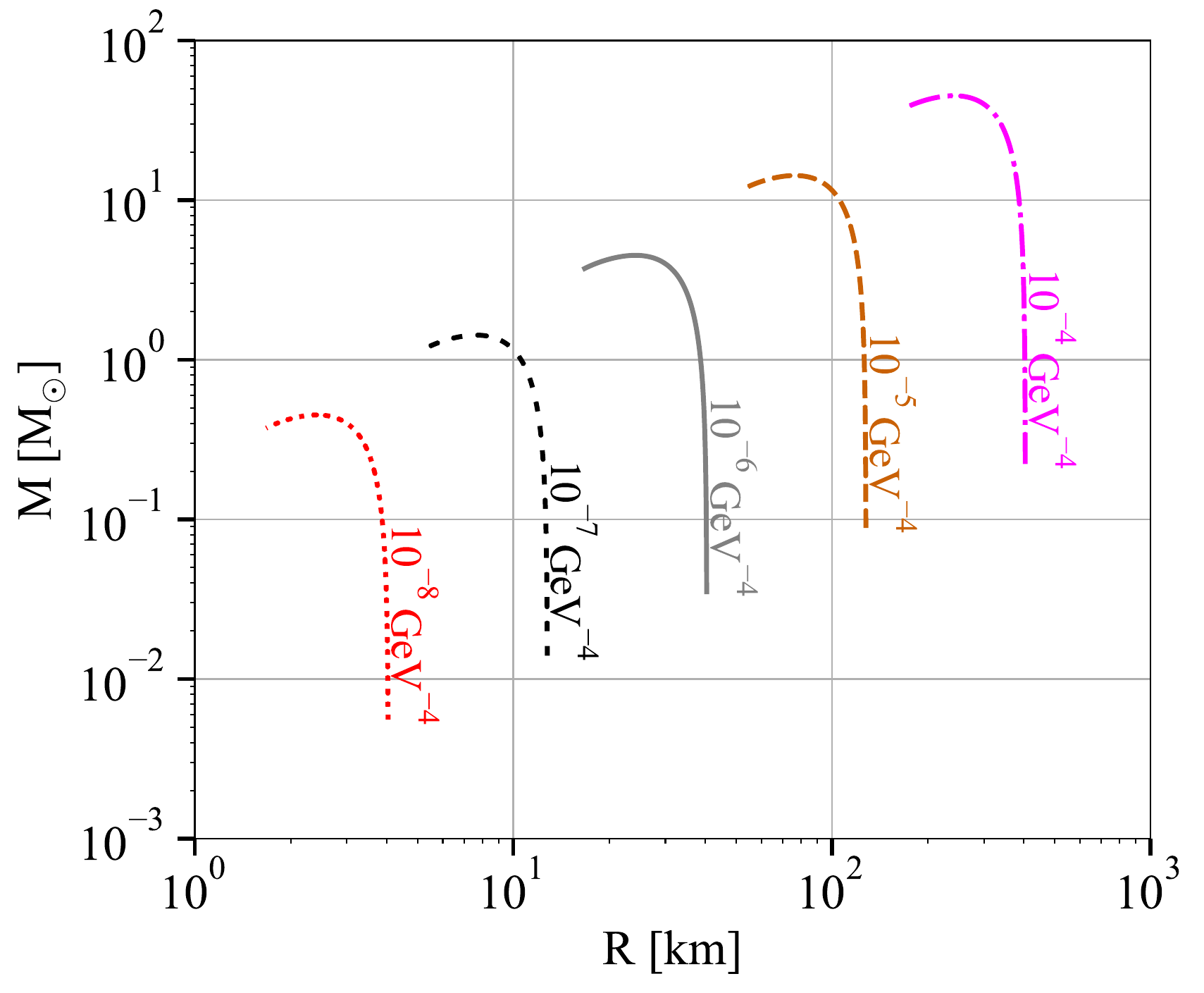}
    \caption{Mass - radius stability curves for boson stars with different ratios involving the self-interaction cross section and mass $\sqrt{\sigma}/m^3$, as labelled.}
    \label{fig:boson star mass radius}
\end{figure}

An important consideration for Figs.~\ref{fig:stability_curves_varied_coupling} and \ref{fig:stability_curves_varied_masspop} is the definition of the radius of neutron star. The presence of the new physics sub-component allows the neutron star to have an extended atmosphere, and so the "radius" is not well-defined, especially in the scenario where the self-interactions are large and the atmosphere is particularly pronounced. Gravitational wave experiments do not distinguish between different types of matter, and the tidal effects depend on the compactness of the object. We define the radius as that which encloses 99.99 percent of the \emph{total} amount of matter in the neutron star, and find this converges sufficiently well.
 
Figure~\ref{fig:boson star mass radius} shows our mass-radius stability curves for a simpler system; boson stars with varying ratios of repulsive self-interactions and mediator masses, with an EOS given by Eq.~\eqref{eq:boson star EOS}. As the boson star is purely new particles (and only is one fluid), its total size is completely dependent on the self-interaction size and particle mass, and can therefore take a significant range of masses and radii. Compared to the neutron star scenario, where the existence of the heaviest neutron star constrains the mass of the object, the boson star mass instead will be constrained by microlensing especially if it becomes too massive, see for example Ref.~\cite{Croon:2020wpr}. In addition, the boson star cannot be arbitrarily massive; if it is held up only by its self-interactions with coupling $\lambda$, then its maximum mass is given by~\cite{PhysRevLett.57.2485}
\begin{equation}
    \textrm{M}_{\rm max} \approx \,\sqrt{\lambda} \left(\frac{ 100\, {\rm MeV}}{m_\chi} \right)^2\, \textrm{M}_\odot.
\end{equation}

We focus on boson stars with radii $ < 10^3 $ km such that the merger frequency falls within the detection window of ground-based interferometer experiments. For such boson stars, tidal forces will affect the binary inspiral in a way approximately similar to NS \cite{Sennett:2017etc}.

\begin{figure*}
    \begin{subfigure}{0.49\textwidth}
    \includegraphics[width=\textwidth]{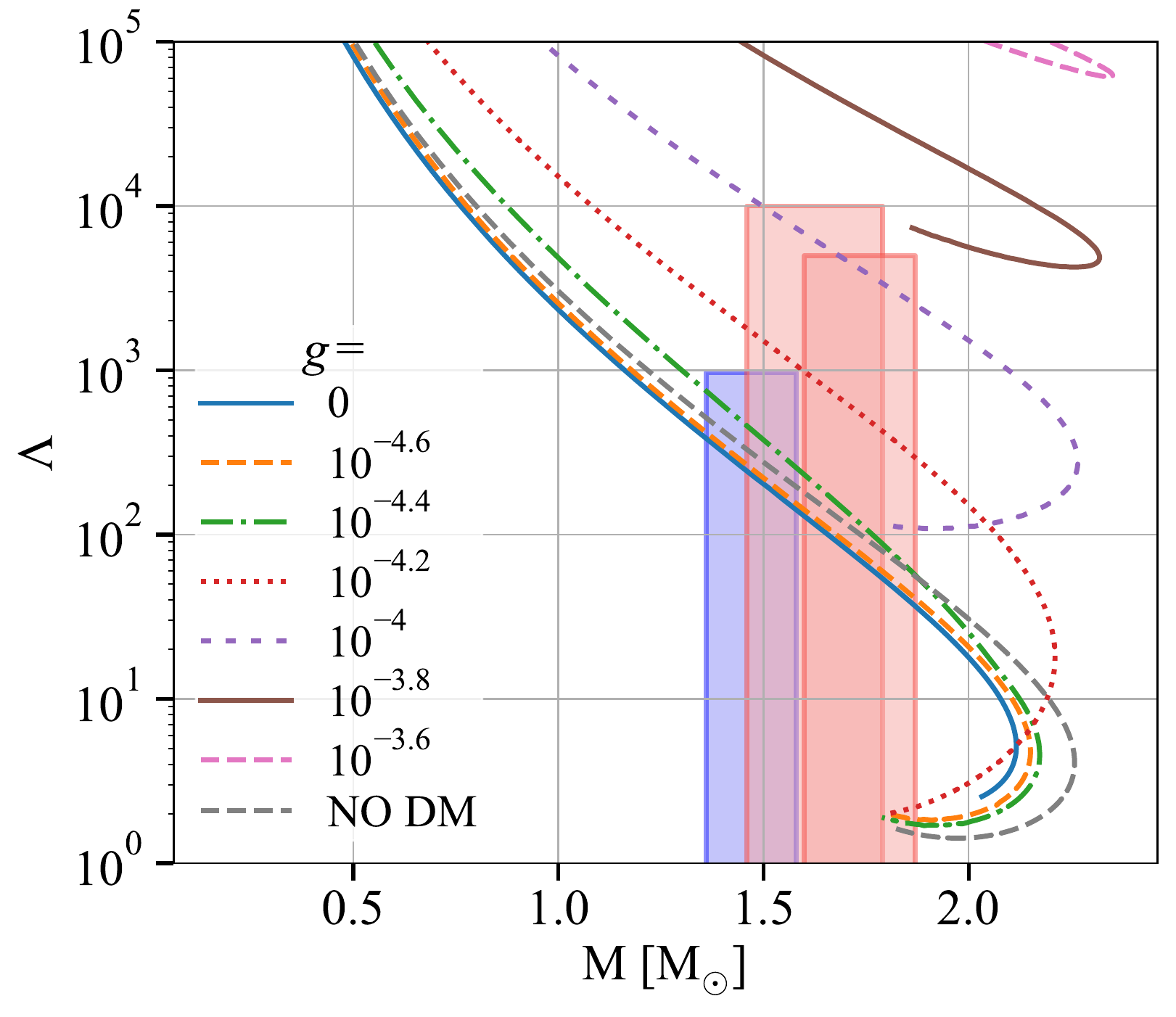}
    \end{subfigure}
\hfill
    \begin{subfigure}{0.49\textwidth}
    \includegraphics[width=\textwidth]{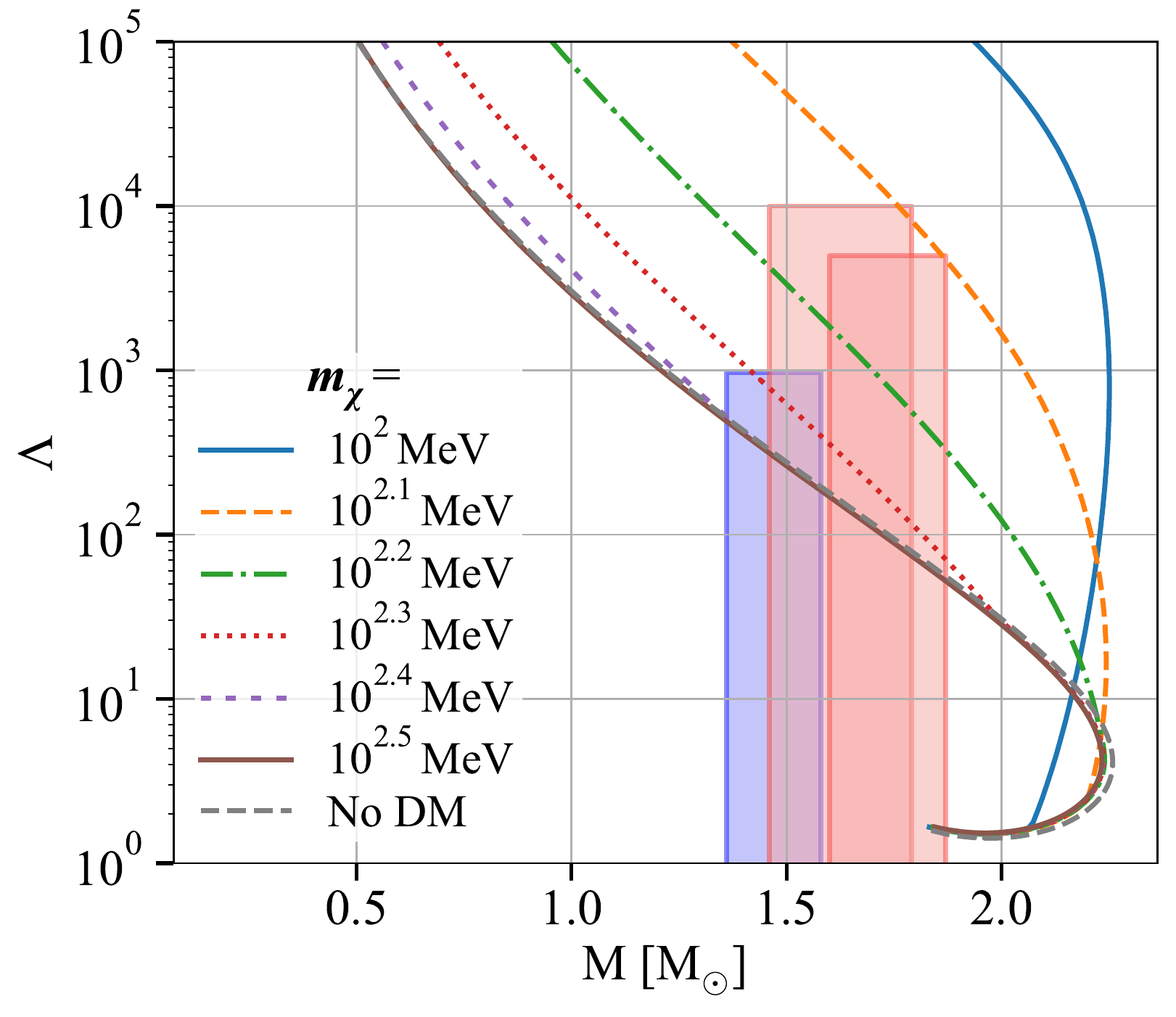}
    \end{subfigure}
\hfill
\vspace{5mm}
   \begin{subfigure}{0.49\textwidth}
    \includegraphics[width=\textwidth]{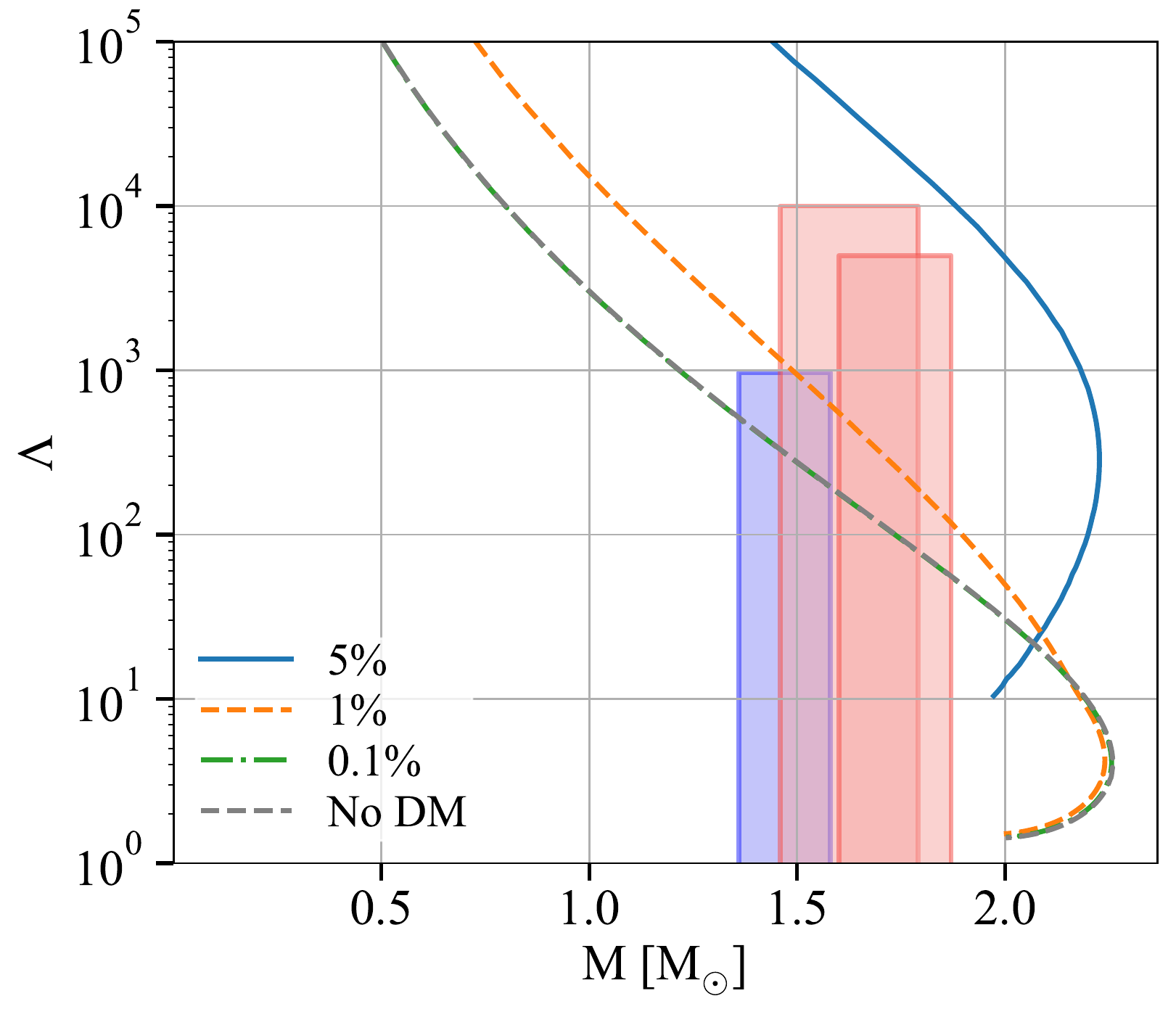}
    \end{subfigure}
    \caption{Tidal deformability-mass stability curves for varying parameters, compared with the 90\% confidence upper bounds on $\Lambda$ for GW170817~\cite{PhysRevLett.119.161101} (blue box) and GW190425~\cite{Abbott_2020} (two red boxes for the primary and secondary components) and the 90\% confidence on the masses of the observed neutron stars \textbf{Top Left:} Impact of varied coupling as labeled, assuming 5\% DM by mass, $m_\chi = 1 \mathrm{\, GeV}$, $m_\phi = 1 \mathrm{\, keV}$. \textbf{Top Right:} Impact of varied DM mass as labeled, assuming a 1\% DM mass fraction with $g = 0$ coupling, and $m_\phi = 1 \mathrm{\, keV}$. \textbf{Bottom:} Impact of varying the DM mass fraction inside the NS, with mass fractions as labeled, and assuming $g = 10^{-5}$  coupling, particle mass $m_\chi = 200 \, \mathrm{MeV}$, and mediator mass $m_\phi = 1 \mathrm{\, keV}$.}
    \label{fig:tidalloveNS}
\end{figure*}

\section{Gravitational wave constraints on celestial objects}
\label{sec:love}

As we saw in the previous section, a range of masses and self-couplings can lead to a new particle atmosphere extended well outside of nuclear matter of a neutron star. For simpler systems such as boson stars, we saw that their radius was also completely dependent on the self-coupling size and particle mass. We now consider the opportunities to probe these macroscopic parameters using gravitational waves.

\subsection{Tidal Love Number Computation}
 The altered radii and atmospheres of celestial objects due to new physics may be probed via tidal interactions. These interactions imprint on the gravitational wave signal of binary neutron star mergers or more hypothetical boson star mergers, in the form of a phase shift, which is given by~\cite{Flanagan:2007ix}
 \begin{equation}
     \delta\Psi=-\frac{117}{256}v^5\frac{M}{\mu}~\tilde{\Lambda}\,,
 \end{equation}
where $\mu$ is the reduced mass, $v=(\pi M f)^{1/3}$ the orbital velocity, $M$ the mass, and
\begin{equation}
    \tilde{\Lambda}=\frac{16}{13} \frac{(M_1+12M_2) M_1^4 \Lambda_1+(12M_1+M_2 )M^4_2\Lambda_2}{(M_1+M_2)^5}\,,
\end{equation}
is the dimensionless measure of the tidal deformablitity in the stellar merger. Here $M_{1,2}$ and $\Lambda_{1,2}$ are the masses and tidal deformabilities of admixed stars. For one component stars (such as boson stars), there is of course only one contribution.

The dimensionful tidal parameter quadrupole is defined as
\begin{equation}
    \lambda=\frac{2}{3}k_2 \left( \frac{GM}{R}\right)^{-5},
\end{equation}
where $k_2$ is the $l=2$ tidal Love number. We use the calculation of $k_2$ is as performed in Ref.~\cite{Hinderer:2007mb}, where it is shown that
\begin{multline}
k_2=\frac{8 C^5}{5} (1 - 2 C)^2 \left(2 (1 - C) + (2 C-1) y_R\right)\times \\
\Bigg\{4 C^3 \left(13 - 11 y_R + 2 C^2 (1 + y_R) + C (-2 + 3 y_R)\right)\\
+2 C \left(6 - 3 y_R + 3 C (5 y_R-8)\right)+3 (1 - 2 C)^2  \\
\times  \left(2 + 2 C (y_R-1) - y_R\right) \ln(1 - 2 C) \Bigg\}^{-1}\, ,
\label{eq:k2}
\end{multline}
where $C$ is the celestial-body compactness, defined as the mass/radius ratio of the celestial body. The parameter $y_R$ above is defined as $y_R=y(R)$, found by solving
\begin{align}
\frac{dy(r)}{dr}=&-\frac{y(r)^2}{r}-\frac{y(r) g_{rr}(r)}{r}\left(1+4\pi r^2(p(r)-\epsilon(r))\right) \nonumber \\
&-4\pi r  \left(9p(r)+5\epsilon(r)+ \frac{d\epsilon}{d p}(p(r)+\epsilon(r))\right)g_{rr}(r) \nonumber \\
&+ r \left(\frac{6g_{rr}(r)}{r^2}+ \left(\frac{d \ln g_{tt}(r)}{dr}\right)^2 \right) \,,
\end{align}
where $g_{rr}(r)$ and $g_{tt}(r)$ are the radial and temporal components of the unperturbed Schwarzschild metric, and $p(r)$ and $\epsilon(r)$ are determined by solving the coupled TOV equations, which are described in the previous section.  Note also that the crust of the NS can impact the tidal deformability~\cite{PhysRevC.101.015806, PhysRevLett.120.261103}.

Note that the tidal deformability parameter $\Lambda$ appearing in waveform models is not always the same as that found in theoretical calculations. Simpler, yet ambiguous, estimates for $\Lambda$ are required due to difficulties in calculating the fifth-order post-Newtonian dynamics; see Ref.~\cite{Gralla:2017djj} for discussion.

\subsection{Admixed Neutron Star Sensitivity}

Figure~\ref{fig:tidalloveNS} shows the tidal deformability and mass stability curves for admixed neutron stars, with a range of parameters, compared with the detected gravitational wave events GW170817 and GW190425 (interpreting the latter as a BNS inspiral; for an alternative interpretation, see \cite{Clesse:2020ghq}). It has also been pointed out that it may be possible to set a lower bound on the tidal deformability, by combining the electromagnetic counterpart of GW170817 as expected from kilonova models with numerical relativity results~\cite{Radice:2017lry}. We do not show this tentative bound in our figure, though it would correspond to $\Lambda\gtrsim400$.

In the top-left panel we show the impact of varying the self-interaction couplings through the range $0-10^{-3.6}$, for a fixed new particle mass of 1 GeV, and a total new particle mass of $5\%$ of the neutron star mass. We see that this measurement is very sensitive to the self-interaction coupling; it must be smaller than about $10^{-4}$ at this mass fraction to not be excluded. This is consistent with the results found in Ref.~\cite{Nelson:2018xtr}.

In the top-right panel of Fig.~\ref{fig:tidalloveNS}, we instead show the impact of varying the new particle mass on the tidal love number, and fix the self-interaction coupling to be zero. For a mass fraction of $1\%$ and light mediator, new particles lighter than around 100 MeV can already be readily tested with gravitational waves, even without any self-interactions. Taken together with the top-left figure, clearly adding self-interactions can lead to even stronger constraints on the new particle sector.

In the bottom panel of Fig.~\ref{fig:tidalloveNS}, we investigate the sensitivity to the total mass fraction of new particles. Here we fix the self-coupling to be small (only $10^{-5}$), and show a MeV-scale (200 MeV) DM/new particle mass. We observe that a range of mass fractions can be constrained by this measurement. When decreasing the DM mass below the 200 MeV mass shown, we find increasing sensitivity to smaller and smaller mass fractions, while increasing the DM mass requires larger mass fractions to be constraining. Overall we find that to constrain a 5 percent new particle/DM mass fraction or less, the DM particle needs to be less than a few hundred MeV in mass with our benchmark light mediator example. As this assumed the small self-coupling of $10^{-5}$, taken together with the other panels of Fig.~\ref{fig:tidalloveNS}, evidently this constraint increases even further for larger self-coupling. Overall, across all these panels, it is clear that hidden new particle sectors are highly testable through NS merger events. Note that our benchmark parameters are consistent with bounds from the Bullet Cluster. A full comparison of the parameter space against Bullet Cluster bounds is shown in Ref.~\cite{Nelson:2018xtr}.

Note that direct observations of NS masses and radii from the NICER x-ray telescope provide complementary access to the NS radius independent of tidal deformability constraints from GW observation of binary neutron star mergers, see Refs.~\cite{Riley_2019,2019ApJ...887L..24M,Miller_2021}.

\subsection{Boson Star Sensitivity}

\begin{figure}[t!]
    \centering
    \includegraphics[scale=0.55]{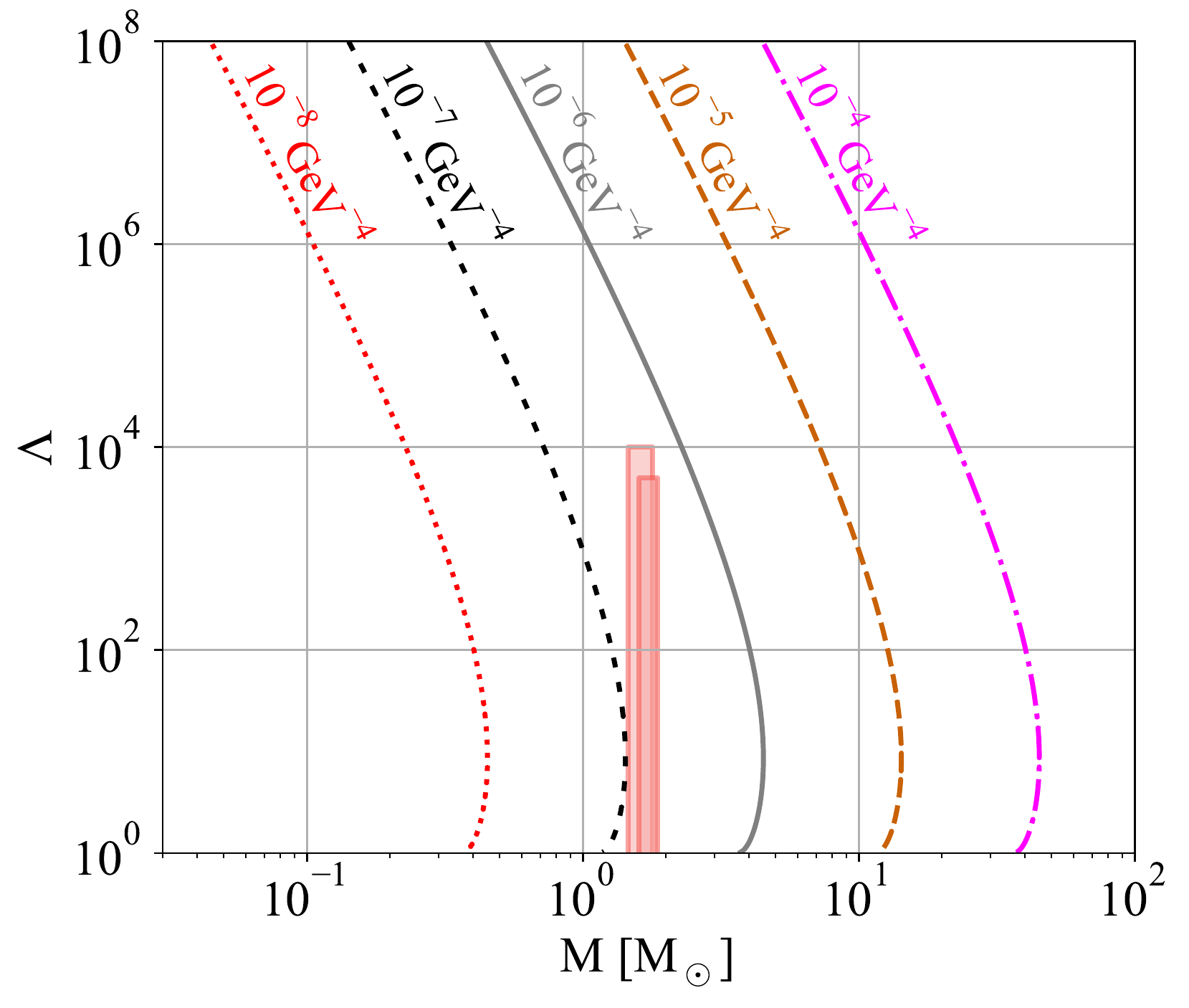}
    \caption{Tidal deformability - mass stability curve for a boson star with EOS given by Eq.~\eqref{eq:boson star EOS} with different ratios involving the self-interaction cross section and mass $\sqrt{\sigma}/m^3$, as labelled. The red boxes show measurements on the primary and secondary objects in the merger event GW190425~\cite{Abbott_2020}.
    }
    \label{fig:boson star deformability}
\end{figure}

Figure~\ref{fig:boson star deformability} shows the tidal love number and mass stability curves for a boson star, corresponding to the setup in Fig.~\ref{fig:boson star mass radius}.  We also show the gravitational wave event GW190425~\cite{Abbott_2020}, which was not known to have had an electromagnetic counterpart. As such, it is possible that this event was a boson star merger, and can be directly compared. Going forward, this means that DM/new particles and self-interactions can be strongly constrained through non-observation or limited observation of boson-star merger events \cite{Croon:2022tmr}. Note that since the boson star EOS only depends on the specific combination $ \sigma/m^3$, constraints on the bosonic self-coupling can only be drawn as a function of the bosonic mass.

\section{Example Applications}
\label{sec:appl}

In the case of boson stars, new particles can simply coalesce into one object, which then may be detectable as we discussed in the previous section. For admixed neutron stars, it is important to note that accumulation of e.g. DM from the Galactic halo leads to negligible DM mass fractions in neutron stars, and so other mechanisms must be considered. In fact, there are a variety of possible scenarios to produce the large abundances required to produce detectable gravitational wave signatures. For completeness we now briefly review and discuss some mechanisms which can produce large quantities of new particles or dark matter inside these objects.

\subsection{Production in Supernovae}

At its birth during a supernova event, a neutron star is very hot, and so can efficiently produce new particles. If these particles are sufficiently light and weakly interacting, they can escape the proto-neutron star (PNS), but they must not exceed the luminosity carried away by neutrinos from within the PNS to the outside of the neutrinosphere~\cite{Raffelt:1996wa}. For the well-studied SN1987A, observations constrain this luminosity to be approximately less than $3\times10^{52}$ erg/s over 10 seconds. Converting this luminosity to a new maximum particle mass, we find that the new mass is $M_\chi\lesssim0.15~$M$_\odot$, such that if all the new particles produced in the PNS are trapped, this is approximately their largest population size. However, any trapped population must not largely annihilate away to have an appreciable effect on the neutron star. As such, some of this mass may still be lost through the new particle trapping mechanism; we give two brief examples in this subsection.

One scenario is that the particles and antiparticles can be pair produced through radiation of a mediator, with an energy splitting, such that a large fraction of (anti)particles could be expelled while the particles with opposite charge remain. This energy splitting could for example result from an attractive (repulsive) interaction between $\chi$ ($\bar\chi$) and nuclear matter mediated by $\phi$; this is the scenario proposed in Ref.~\cite{Nelson:2018xtr}. This setup produces a nearly fully trapped new particle population, and nearly full expelled anti-particle population, for example parameters of $g_\chi\sim1$, $g_B\sim10^{-9.8}$, $m_\chi\sim50$ MeV, and $m_\phi\sim1$ keV, which produces a total DM mass of approximately $\sim0.08\,M_\odot$ (approximately half of the total possible mass). These large self-interaction sizes are clearly testable as shown in Fig.~\ref{fig:tidalloveNS}.

An alternate scenario to consider is a particle model which leads to a non-annihilating population of particles which are fully trapped, and not expelled. This can be realized in the context of an inelastic model, with a mass splitting between two particle states (this class of models is often investigated in the context of "inelastic dark matter" models). If the attractive force is too weak, both the DM states may escape the PNS. If the attractive force is sufficiently strong, all of the lighter and heavier DM states are trapped in the PNS. In an intermediate regime, some of both the lighter and heavier DM remains trapped and some escapes, with the difference driven by their differing gravitational potentials, rather than energy from charge potential in the example discussed above. When both states are trapped in the PNS, it is possible that only the lighter state dominantly remains, as the heavier state rapidly decays into the lighter state. While the lighter state may annihilate to mediator pairs at tree level through a $t$-channel exchange of the heavier new particle, for a maximally CP violating model, such interactions may be forbidden. Lastly, while elastic new particle annihilation can occur at loop-level, this process may be very suppressed compared to the tree-level inelastic process, if tiny $\chi$-SM couplings are considered. Note that as the decay of the heavier state into the lighter state plus e.g. electrons can lead to energy being reprocessed into the SM sector, some additional mass may be lost; simulations may be required to determine the precise abundances for such a setup. In any case, such a scenario may provide another way to achieve a large new particle abundance, for MeV-scale new particles with large testable self-interactions as per Fig.~\ref{fig:tidalloveNS}. Overall, we emphasize that simulations would be required to determine precise new particle abundances, and here we instead simply discuss some potential mechanisms.

\subsection{Production in Neutron Decay}

GeV-scale new particles may be trapped in neutron stars after being produced via neutron decay. The Fermi momentum of neutrons inside neutron stars is high, and allows for potential new particle production with masses less than approximately $m_n+\mathcal{O}(p_F/2m_n)$. As discussed in Ref.~\cite{Ellis:2018bkr}, this can produce a few percent levels of new particles in neutron stars.

\subsection{Formation from Dark Stars}

A more hypothetical scenario is where the new particle or dark matter structure is formed independent of the neutron star, and at a later stage become intertwined with it. This may occur as dissipation within the dark sector could lead to dark clumps, which may either seed the formation of a star, or merge with a companion star~\cite{Ciarcelluti:2010ji}.

\section{Summary and Conclusions}
\label{sec:conclusion}
The coming decade of gravitational wave astronomy will open the door to compact object spectroscopy using their tidal interactions in binary mergers. In this work, we have described a generic framework to solve TOV equations for two fluids, applicable to arbitrary celestial objects and varying new particle/DM population sizes and properties.
Our work is accompanied by a publicly released \verb_Python_ code with which the tidal love number for two arbitrary fluids with a given equation of state and mass fraction can be calculated, available at \href{https://zenodo.org/record/7361819\#.Y4DebaLP3JE}{https://zenodo.org/record/7361819\#.Y4DebaLP3JE}.

In this work, we have applied this framework to two situations of common interest: neutron stars admixed with a dark fluid, and Bose-Einstein condensates from a boson with a repulsive self-interaction. In the former case, our results support and generalise the conclusions drawn in Ref.~\cite{Nelson:2018xtr} using GW170817: for a given mass fraction of the dark fluid, strong constraints can be derived on the hidden sector mass and gauge coupling. Future binary neutron star observations can be straightforwardly compared to the predictions shown in Fig.~\ref{fig:tidalloveNS}.

For boson stars consisting on a single particle with repulsive self-interactions, the tidal love number depends on the mass of the boson and the self-interaction through the combination $ \sqrt{\sigma}/m^3$, as shown in Fig.~\ref{fig:boson star deformability}. As the more recent GW190425 was not known to have had an electromagnetic counterpart, we showed that new particles and self-interactions can also be strongly constrained through the lack of boson-star merger events. The non-observation (or limited observation) of boson star mergers can also be interpreted as a constraint on the dark matter mass fraction such objects can comprise, which is explored in future work~\cite{Croon:2022tmr}.

\section*{Acknowledgments}

We thank Moira Gresham, Sam McDermott, David McKeen, Sanjay Reddy, Dake Zhou, and Kathryn Zurek, for helpful discussions. DC and RKL thank the Munich Institute for Astro, -Particle and BioPhysics (MIAPbP) for their hospitality and support during completion of this work, which is funded by the Deutsche Forschungsgemeinschaft (DFG, German Research Foundation) under Germany´s Excellence Strategy – EXC-2094 – 390783311. DC is supported by the STFC under Grant No. ST/T001011/1. RKL is supported by the U.S. Department of Energy under Contract DE-AC02-76SF00515.

\bibliography{main}

\begin{thebibliography}{95}%
\makeatletter
\providecommand \@ifxundefined [1]{%
 \@ifx{#1\undefined}
}%
\providecommand \@ifnum [1]{%
 \ifnum #1\expandafter \@firstoftwo
 \else \expandafter \@secondoftwo
 \fi
}%
\providecommand \@ifx [1]{%
 \ifx #1\expandafter \@firstoftwo
 \else \expandafter \@secondoftwo
 \fi
}%
\providecommand \natexlab [1]{#1}%
\providecommand \enquote  [1]{``#1''}%
\providecommand \bibnamefont  [1]{#1}%
\providecommand \bibfnamefont [1]{#1}%
\providecommand \citenamefont [1]{#1}%
\providecommand \href@noop [0]{\@secondoftwo}%
\providecommand \href [0]{\begingroup \@sanitize@url \@href}%
\providecommand \@href[1]{\@@startlink{#1}\@@href}%
\providecommand \@@href[1]{\endgroup#1\@@endlink}%
\providecommand \@sanitize@url [0]{\catcode `\\12\catcode `\$12\catcode
  `\&12\catcode `\#12\catcode `\^12\catcode `\_12\catcode `\%12\relax}%
\providecommand \@@startlink[1]{}%
\providecommand \@@endlink[0]{}%
\providecommand \url  [0]{\begingroup\@sanitize@url \@url }%
\providecommand \@url [1]{\endgroup\@href {#1}{\urlprefix }}%
\providecommand \urlprefix  [0]{URL }%
\providecommand \Eprint [0]{\href }%
\providecommand \doibase [0]{http://dx.doi.org/}%
\providecommand \selectlanguage [0]{\@gobble}%
\providecommand \bibinfo  [0]{\@secondoftwo}%
\providecommand \bibfield  [0]{\@secondoftwo}%
\providecommand \translation [1]{[#1]}%
\providecommand \BibitemOpen [0]{}%
\providecommand \bibitemStop [0]{}%
\providecommand \bibitemNoStop [0]{.\EOS\space}%
\providecommand \EOS [0]{\spacefactor3000\relax}%
\providecommand \BibitemShut  [1]{\csname bibitem#1\endcsname}%
\let\auto@bib@innerbib\@empty
\bibitem [{\citenamefont {Goldman}\ and\ \citenamefont
  {Nussinov}(1989{\natexlab{a}})}]{Goldman:1989nd}%
  \BibitemOpen
  \bibfield  {author} {\bibinfo {author} {\bibfnamefont {I.}~\bibnamefont
  {Goldman}}\ and\ \bibinfo {author} {\bibfnamefont {S.}~\bibnamefont
  {Nussinov}},\ }\href {\doibase 10.1103/PhysRevD.40.3221} {\bibfield
  {journal} {\bibinfo  {journal} {Phys. Rev. D}\ }\textbf {\bibinfo {volume}
  {40}},\ \bibinfo {pages} {3221} (\bibinfo {year}
  {1989}{\natexlab{a}})}\BibitemShut {NoStop}%
\bibitem [{\citenamefont {Bertone}\ and\ \citenamefont
  {Fairbairn}(2008)}]{Bertone:2007ae}%
  \BibitemOpen
  \bibfield  {author} {\bibinfo {author} {\bibfnamefont {G.}~\bibnamefont
  {Bertone}}\ and\ \bibinfo {author} {\bibfnamefont {M.}~\bibnamefont
  {Fairbairn}},\ }\href {\doibase 10.1103/PhysRevD.77.043515} {\bibfield
  {journal} {\bibinfo  {journal} {Phys. Rev. D}\ }\textbf {\bibinfo {volume}
  {77}},\ \bibinfo {pages} {043515} (\bibinfo {year} {2008})},\ \Eprint
  {http://arxiv.org/abs/0709.1485} {arXiv:0709.1485 [astro-ph]} \BibitemShut
  {NoStop}%
\bibitem [{\citenamefont {Mack}\ \emph {et~al.}(2007)\citenamefont {Mack},
  \citenamefont {Beacom},\ and\ \citenamefont {Bertone}}]{Mack:2007xj}%
  \BibitemOpen
  \bibfield  {author} {\bibinfo {author} {\bibfnamefont {G.~D.}\ \bibnamefont
  {Mack}}, \bibinfo {author} {\bibfnamefont {J.~F.}\ \bibnamefont {Beacom}}, \
  and\ \bibinfo {author} {\bibfnamefont {G.}~\bibnamefont {Bertone}},\ }\href
  {\doibase 10.1103/PhysRevD.76.043523} {\bibfield  {journal} {\bibinfo
  {journal} {Phys. Rev. D}\ }\textbf {\bibinfo {volume} {76}},\ \bibinfo
  {pages} {043523} (\bibinfo {year} {2007})},\ \Eprint
  {http://arxiv.org/abs/0705.4298} {arXiv:0705.4298 [astro-ph]} \BibitemShut
  {NoStop}%
\bibitem [{\citenamefont {de~Lavallaz}\ and\ \citenamefont
  {Fairbairn}(2010)}]{deLavallaz:2010wp}%
  \BibitemOpen
  \bibfield  {author} {\bibinfo {author} {\bibfnamefont {A.}~\bibnamefont
  {de~Lavallaz}}\ and\ \bibinfo {author} {\bibfnamefont {M.}~\bibnamefont
  {Fairbairn}},\ }\href {\doibase 10.1103/PhysRevD.81.123521} {\bibfield
  {journal} {\bibinfo  {journal} {Phys. Rev. D}\ }\textbf {\bibinfo {volume}
  {81}},\ \bibinfo {pages} {123521} (\bibinfo {year} {2010})},\ \Eprint
  {http://arxiv.org/abs/1004.0629} {arXiv:1004.0629 [astro-ph.GA]} \BibitemShut
  {NoStop}%
\bibitem [{\citenamefont {Kouvaris}\ and\ \citenamefont
  {Tinyakov}(2010{\natexlab{a}})}]{Kouvaris:2010vv}%
  \BibitemOpen
  \bibfield  {author} {\bibinfo {author} {\bibfnamefont {C.}~\bibnamefont
  {Kouvaris}}\ and\ \bibinfo {author} {\bibfnamefont {P.}~\bibnamefont
  {Tinyakov}},\ }\href {\doibase 10.1103/PhysRevD.82.063531} {\bibfield
  {journal} {\bibinfo  {journal} {Phys. Rev. D}\ }\textbf {\bibinfo {volume}
  {82}},\ \bibinfo {pages} {063531} (\bibinfo {year} {2010}{\natexlab{a}})},\
  \Eprint {http://arxiv.org/abs/1004.0586} {arXiv:1004.0586 [astro-ph.GA]}
  \BibitemShut {NoStop}%
\bibitem [{\citenamefont {McCullough}\ and\ \citenamefont
  {Fairbairn}(2010)}]{McCullough:2010ai}%
  \BibitemOpen
  \bibfield  {author} {\bibinfo {author} {\bibfnamefont {M.}~\bibnamefont
  {McCullough}}\ and\ \bibinfo {author} {\bibfnamefont {M.}~\bibnamefont
  {Fairbairn}},\ }\href {\doibase 10.1103/PhysRevD.81.083520} {\bibfield
  {journal} {\bibinfo  {journal} {Phys. Rev.}\ }\textbf {\bibinfo {volume}
  {D81}},\ \bibinfo {pages} {083520} (\bibinfo {year} {2010})},\ \Eprint
  {http://arxiv.org/abs/1001.2737} {arXiv:1001.2737 [hep-ph]} \BibitemShut
  {NoStop}%
\bibitem [{\citenamefont {Baryakhtar}\ \emph {et~al.}(2017)\citenamefont
  {Baryakhtar}, \citenamefont {Bramante}, \citenamefont {Li}, \citenamefont
  {Linden},\ and\ \citenamefont {Raj}}]{Baryakhtar:2017dbj}%
  \BibitemOpen
  \bibfield  {author} {\bibinfo {author} {\bibfnamefont {M.}~\bibnamefont
  {Baryakhtar}}, \bibinfo {author} {\bibfnamefont {J.}~\bibnamefont
  {Bramante}}, \bibinfo {author} {\bibfnamefont {S.~W.}\ \bibnamefont {Li}},
  \bibinfo {author} {\bibfnamefont {T.}~\bibnamefont {Linden}}, \ and\ \bibinfo
  {author} {\bibfnamefont {N.}~\bibnamefont {Raj}},\ }\href {\doibase
  10.1103/PhysRevLett.119.131801} {\bibfield  {journal} {\bibinfo  {journal}
  {Phys. Rev. Lett.}\ }\textbf {\bibinfo {volume} {119}},\ \bibinfo {pages}
  {131801} (\bibinfo {year} {2017})},\ \Eprint
  {http://arxiv.org/abs/1704.01577} {arXiv:1704.01577 [hep-ph]} \BibitemShut
  {NoStop}%
\bibitem [{\citenamefont {Raj}\ \emph {et~al.}(2018)\citenamefont {Raj},
  \citenamefont {Tanedo},\ and\ \citenamefont {Yu}}]{Raj:2017wrv}%
  \BibitemOpen
  \bibfield  {author} {\bibinfo {author} {\bibfnamefont {N.}~\bibnamefont
  {Raj}}, \bibinfo {author} {\bibfnamefont {P.}~\bibnamefont {Tanedo}}, \ and\
  \bibinfo {author} {\bibfnamefont {H.-B.}\ \bibnamefont {Yu}},\ }\href
  {\doibase 10.1103/PhysRevD.97.043006} {\bibfield  {journal} {\bibinfo
  {journal} {Phys. Rev.}\ }\textbf {\bibinfo {volume} {D97}},\ \bibinfo {pages}
  {043006} (\bibinfo {year} {2018})},\ \Eprint
  {http://arxiv.org/abs/1707.09442} {arXiv:1707.09442 [hep-ph]} \BibitemShut
  {NoStop}%
\bibitem [{\citenamefont {Bell}\ \emph {et~al.}(2018)\citenamefont {Bell},
  \citenamefont {Busoni},\ and\ \citenamefont {Robles}}]{Bell:2018pkk}%
  \BibitemOpen
  \bibfield  {author} {\bibinfo {author} {\bibfnamefont {N.~F.}\ \bibnamefont
  {Bell}}, \bibinfo {author} {\bibfnamefont {G.}~\bibnamefont {Busoni}}, \ and\
  \bibinfo {author} {\bibfnamefont {S.}~\bibnamefont {Robles}},\ }\href
  {\doibase 10.1088/1475-7516/2018/09/018} {\bibfield  {journal} {\bibinfo
  {journal} {JCAP}\ }\textbf {\bibinfo {volume} {1809}},\ \bibinfo {pages}
  {018} (\bibinfo {year} {2018})},\ \Eprint {http://arxiv.org/abs/1807.02840}
  {arXiv:1807.02840 [hep-ph]} \BibitemShut {NoStop}%
\bibitem [{\citenamefont {Chen}\ and\ \citenamefont
  {Lin}(2018)}]{Chen:2018ohx}%
  \BibitemOpen
  \bibfield  {author} {\bibinfo {author} {\bibfnamefont {C.-S.}\ \bibnamefont
  {Chen}}\ and\ \bibinfo {author} {\bibfnamefont {Y.-H.}\ \bibnamefont {Lin}},\
  }\href {\doibase 10.1007/JHEP08(2018)069} {\bibfield  {journal} {\bibinfo
  {journal} {JHEP}\ }\textbf {\bibinfo {volume} {08}},\ \bibinfo {pages} {069}
  (\bibinfo {year} {2018})},\ \Eprint {http://arxiv.org/abs/1804.03409}
  {arXiv:1804.03409 [hep-ph]} \BibitemShut {NoStop}%
\bibitem [{\citenamefont {Dasgupta}\ \emph {et~al.}(2019)\citenamefont
  {Dasgupta}, \citenamefont {Gupta},\ and\ \citenamefont
  {Ray}}]{Dasgupta:2019juq}%
  \BibitemOpen
  \bibfield  {author} {\bibinfo {author} {\bibfnamefont {B.}~\bibnamefont
  {Dasgupta}}, \bibinfo {author} {\bibfnamefont {A.}~\bibnamefont {Gupta}}, \
  and\ \bibinfo {author} {\bibfnamefont {A.}~\bibnamefont {Ray}},\ }\href
  {\doibase 10.1088/1475-7516/2019/08/018} {\bibfield  {journal} {\bibinfo
  {journal} {JCAP}\ }\textbf {\bibinfo {volume} {08}},\ \bibinfo {pages} {018}
  (\bibinfo {year} {2019})},\ \Eprint {http://arxiv.org/abs/1906.04204}
  {arXiv:1906.04204 [hep-ph]} \BibitemShut {NoStop}%
\bibitem [{\citenamefont {Hamaguchi}\ \emph {et~al.}(2019)\citenamefont
  {Hamaguchi}, \citenamefont {Nagata},\ and\ \citenamefont
  {Yanagi}}]{Hamaguchi:2019oev}%
  \BibitemOpen
  \bibfield  {author} {\bibinfo {author} {\bibfnamefont {K.}~\bibnamefont
  {Hamaguchi}}, \bibinfo {author} {\bibfnamefont {N.}~\bibnamefont {Nagata}}, \
  and\ \bibinfo {author} {\bibfnamefont {K.}~\bibnamefont {Yanagi}},\ }\href
  {\doibase 10.1016/j.physletb.2019.06.060} {\bibfield  {journal} {\bibinfo
  {journal} {Phys. Lett.}\ }\textbf {\bibinfo {volume} {B795}},\ \bibinfo
  {pages} {484} (\bibinfo {year} {2019})},\ \Eprint
  {http://arxiv.org/abs/1905.02991} {arXiv:1905.02991 [hep-ph]} \BibitemShut
  {NoStop}%
\bibitem [{\citenamefont {Camargo}\ \emph {et~al.}(2019)\citenamefont
  {Camargo}, \citenamefont {Queiroz},\ and\ \citenamefont
  {Sturani}}]{Camargo:2019wou}%
  \BibitemOpen
  \bibfield  {author} {\bibinfo {author} {\bibfnamefont {D.~A.}\ \bibnamefont
  {Camargo}}, \bibinfo {author} {\bibfnamefont {F.~S.}\ \bibnamefont
  {Queiroz}}, \ and\ \bibinfo {author} {\bibfnamefont {R.}~\bibnamefont
  {Sturani}},\ }\href {\doibase 10.1088/1475-7516/2019/09/051} {\bibfield
  {journal} {\bibinfo  {journal} {JCAP}\ }\textbf {\bibinfo {volume} {1909}},\
  \bibinfo {pages} {051} (\bibinfo {year} {2019})},\ \Eprint
  {http://arxiv.org/abs/1901.05474} {arXiv:1901.05474 [hep-ph]} \BibitemShut
  {NoStop}%
\bibitem [{\citenamefont {Bell}\ \emph {et~al.}(2019)\citenamefont {Bell},
  \citenamefont {Busoni},\ and\ \citenamefont {Robles}}]{Bell:2019pyc}%
  \BibitemOpen
  \bibfield  {author} {\bibinfo {author} {\bibfnamefont {N.~F.}\ \bibnamefont
  {Bell}}, \bibinfo {author} {\bibfnamefont {G.}~\bibnamefont {Busoni}}, \ and\
  \bibinfo {author} {\bibfnamefont {S.}~\bibnamefont {Robles}},\ }\href
  {\doibase 10.1088/1475-7516/2019/06/054} {\bibfield  {journal} {\bibinfo
  {journal} {JCAP}\ }\textbf {\bibinfo {volume} {1906}},\ \bibinfo {pages}
  {054} (\bibinfo {year} {2019})},\ \Eprint {http://arxiv.org/abs/1904.09803}
  {arXiv:1904.09803 [hep-ph]} \BibitemShut {NoStop}%
\bibitem [{\citenamefont {Acevedo}\ \emph {et~al.}(2020)\citenamefont
  {Acevedo}, \citenamefont {Bramante}, \citenamefont {Leane},\ and\
  \citenamefont {Raj}}]{Acevedo:2019agu}%
  \BibitemOpen
  \bibfield  {author} {\bibinfo {author} {\bibfnamefont {J.~F.}\ \bibnamefont
  {Acevedo}}, \bibinfo {author} {\bibfnamefont {J.}~\bibnamefont {Bramante}},
  \bibinfo {author} {\bibfnamefont {R.~K.}\ \bibnamefont {Leane}}, \ and\
  \bibinfo {author} {\bibfnamefont {N.}~\bibnamefont {Raj}},\ }\href {\doibase
  10.1088/1475-7516/2020/03/038} {\bibfield  {journal} {\bibinfo  {journal}
  {JCAP}\ }\textbf {\bibinfo {volume} {03}},\ \bibinfo {pages} {038} (\bibinfo
  {year} {2020})},\ \Eprint {http://arxiv.org/abs/1911.06334} {arXiv:1911.06334
  [hep-ph]} \BibitemShut {NoStop}%
\bibitem [{\citenamefont {Joglekar}\ \emph {et~al.}(2019)\citenamefont
  {Joglekar}, \citenamefont {Raj}, \citenamefont {Tanedo},\ and\ \citenamefont
  {Yu}}]{Joglekar:2019vzy}%
  \BibitemOpen
  \bibfield  {author} {\bibinfo {author} {\bibfnamefont {A.}~\bibnamefont
  {Joglekar}}, \bibinfo {author} {\bibfnamefont {N.}~\bibnamefont {Raj}},
  \bibinfo {author} {\bibfnamefont {P.}~\bibnamefont {Tanedo}}, \ and\ \bibinfo
  {author} {\bibfnamefont {H.-B.}\ \bibnamefont {Yu}},\ }\href@noop {} {\
  (\bibinfo {year} {2019})},\ \Eprint {http://arxiv.org/abs/1911.13293}
  {arXiv:1911.13293 [hep-ph]} \BibitemShut {NoStop}%
\bibitem [{\citenamefont {Joglekar}\ \emph {et~al.}(2020)\citenamefont
  {Joglekar}, \citenamefont {Raj}, \citenamefont {Tanedo},\ and\ \citenamefont
  {Yu}}]{Joglekar:2020liw}%
  \BibitemOpen
  \bibfield  {author} {\bibinfo {author} {\bibfnamefont {A.}~\bibnamefont
  {Joglekar}}, \bibinfo {author} {\bibfnamefont {N.}~\bibnamefont {Raj}},
  \bibinfo {author} {\bibfnamefont {P.}~\bibnamefont {Tanedo}}, \ and\ \bibinfo
  {author} {\bibfnamefont {H.-B.}\ \bibnamefont {Yu}},\ }\href@noop {} {\
  (\bibinfo {year} {2020})},\ \Eprint {http://arxiv.org/abs/2004.09539}
  {arXiv:2004.09539 [hep-ph]} \BibitemShut {NoStop}%
\bibitem [{\citenamefont {Leane}\ and\ \citenamefont
  {Smirnov}(2021)}]{Leane:2020wob}%
  \BibitemOpen
  \bibfield  {author} {\bibinfo {author} {\bibfnamefont {R.~K.}\ \bibnamefont
  {Leane}}\ and\ \bibinfo {author} {\bibfnamefont {J.}~\bibnamefont
  {Smirnov}},\ }\href {\doibase 10.1103/PhysRevLett.126.161101} {\bibfield
  {journal} {\bibinfo  {journal} {Phys. Rev. Lett.}\ }\textbf {\bibinfo
  {volume} {126}},\ \bibinfo {pages} {161101} (\bibinfo {year} {2021})},\
  \Eprint {http://arxiv.org/abs/2010.00015} {arXiv:2010.00015 [hep-ph]}
  \BibitemShut {NoStop}%
\bibitem [{\citenamefont {Bell}\ \emph {et~al.}(2020)\citenamefont {Bell},
  \citenamefont {Busoni}, \citenamefont {Robles},\ and\ \citenamefont
  {Virgato}}]{Bell:2020jou}%
  \BibitemOpen
  \bibfield  {author} {\bibinfo {author} {\bibfnamefont {N.~F.}\ \bibnamefont
  {Bell}}, \bibinfo {author} {\bibfnamefont {G.}~\bibnamefont {Busoni}},
  \bibinfo {author} {\bibfnamefont {S.}~\bibnamefont {Robles}}, \ and\ \bibinfo
  {author} {\bibfnamefont {M.}~\bibnamefont {Virgato}},\ }\href@noop {} {\
  (\bibinfo {year} {2020})},\ \Eprint {http://arxiv.org/abs/2004.14888}
  {arXiv:2004.14888 [hep-ph]} \BibitemShut {NoStop}%
\bibitem [{\citenamefont {Dasgupta}\ \emph {et~al.}(2020)\citenamefont
  {Dasgupta}, \citenamefont {Gupta},\ and\ \citenamefont
  {Ray}}]{Dasgupta:2020dik}%
  \BibitemOpen
  \bibfield  {author} {\bibinfo {author} {\bibfnamefont {B.}~\bibnamefont
  {Dasgupta}}, \bibinfo {author} {\bibfnamefont {A.}~\bibnamefont {Gupta}}, \
  and\ \bibinfo {author} {\bibfnamefont {A.}~\bibnamefont {Ray}},\ }\href@noop
  {} {\  (\bibinfo {year} {2020})},\ \Eprint {http://arxiv.org/abs/2006.10773}
  {arXiv:2006.10773 [hep-ph]} \BibitemShut {NoStop}%
\bibitem [{\citenamefont {Garani}\ \emph {et~al.}(2020)\citenamefont {Garani},
  \citenamefont {Gupta},\ and\ \citenamefont {Raj}}]{Garani:2020wge}%
  \BibitemOpen
  \bibfield  {author} {\bibinfo {author} {\bibfnamefont {R.}~\bibnamefont
  {Garani}}, \bibinfo {author} {\bibfnamefont {A.}~\bibnamefont {Gupta}}, \
  and\ \bibinfo {author} {\bibfnamefont {N.}~\bibnamefont {Raj}},\ }\href@noop
  {} {\  (\bibinfo {year} {2020})},\ \Eprint {http://arxiv.org/abs/2009.10728}
  {arXiv:2009.10728 [hep-ph]} \BibitemShut {NoStop}%
\bibitem [{\citenamefont {Bramante}\ \emph {et~al.}(2020)\citenamefont
  {Bramante}, \citenamefont {Buchanan}, \citenamefont {Goodman},\ and\
  \citenamefont {Lodhi}}]{Bramante:2019fhi}%
  \BibitemOpen
  \bibfield  {author} {\bibinfo {author} {\bibfnamefont {J.}~\bibnamefont
  {Bramante}}, \bibinfo {author} {\bibfnamefont {A.}~\bibnamefont {Buchanan}},
  \bibinfo {author} {\bibfnamefont {A.}~\bibnamefont {Goodman}}, \ and\
  \bibinfo {author} {\bibfnamefont {E.}~\bibnamefont {Lodhi}},\ }\href
  {\doibase 10.1103/PhysRevD.101.043001} {\bibfield  {journal} {\bibinfo
  {journal} {Phys. Rev. D}\ }\textbf {\bibinfo {volume} {101}},\ \bibinfo
  {pages} {043001} (\bibinfo {year} {2020})},\ \Eprint
  {http://arxiv.org/abs/1909.11683} {arXiv:1909.11683 [hep-ph]} \BibitemShut
  {NoStop}%
\bibitem [{\citenamefont {Bell}\ \emph {et~al.}(2021)\citenamefont {Bell},
  \citenamefont {Busoni}, \citenamefont {Ramirez-Quezada}, \citenamefont
  {Robles},\ and\ \citenamefont {Virgato}}]{Bell:2021fye}%
  \BibitemOpen
  \bibfield  {author} {\bibinfo {author} {\bibfnamefont {N.~F.}\ \bibnamefont
  {Bell}}, \bibinfo {author} {\bibfnamefont {G.}~\bibnamefont {Busoni}},
  \bibinfo {author} {\bibfnamefont {M.~E.}\ \bibnamefont {Ramirez-Quezada}},
  \bibinfo {author} {\bibfnamefont {S.}~\bibnamefont {Robles}}, \ and\ \bibinfo
  {author} {\bibfnamefont {M.}~\bibnamefont {Virgato}},\ }\href {\doibase
  10.1088/1475-7516/2021/10/083} {\bibfield  {journal} {\bibinfo  {journal}
  {JCAP}\ }\textbf {\bibinfo {volume} {10}},\ \bibinfo {pages} {083} (\bibinfo
  {year} {2021})},\ \Eprint {http://arxiv.org/abs/2104.14367} {arXiv:2104.14367
  [hep-ph]} \BibitemShut {NoStop}%
\bibitem [{\citenamefont {Freese}\ \emph {et~al.}(2009)\citenamefont {Freese},
  \citenamefont {Gondolo}, \citenamefont {Sellwood},\ and\ \citenamefont
  {Spolyar}}]{Freese:2008hb}%
  \BibitemOpen
  \bibfield  {author} {\bibinfo {author} {\bibfnamefont {K.}~\bibnamefont
  {Freese}}, \bibinfo {author} {\bibfnamefont {P.}~\bibnamefont {Gondolo}},
  \bibinfo {author} {\bibfnamefont {J.~A.}\ \bibnamefont {Sellwood}}, \ and\
  \bibinfo {author} {\bibfnamefont {D.}~\bibnamefont {Spolyar}},\ }\href
  {\doibase 10.1088/0004-637X/693/2/1563} {\bibfield  {journal} {\bibinfo
  {journal} {Astrophys. J.}\ }\textbf {\bibinfo {volume} {693}},\ \bibinfo
  {pages} {1563} (\bibinfo {year} {2009})},\ \Eprint
  {http://arxiv.org/abs/0805.3540} {arXiv:0805.3540 [astro-ph]} \BibitemShut
  {NoStop}%
\bibitem [{\citenamefont {Taoso}\ \emph {et~al.}(2008)\citenamefont {Taoso},
  \citenamefont {Bertone}, \citenamefont {Meynet},\ and\ \citenamefont
  {Ekstrom}}]{Taoso:2008kw}%
  \BibitemOpen
  \bibfield  {author} {\bibinfo {author} {\bibfnamefont {M.}~\bibnamefont
  {Taoso}}, \bibinfo {author} {\bibfnamefont {G.}~\bibnamefont {Bertone}},
  \bibinfo {author} {\bibfnamefont {G.}~\bibnamefont {Meynet}}, \ and\ \bibinfo
  {author} {\bibfnamefont {S.}~\bibnamefont {Ekstrom}},\ }\href {\doibase
  10.1103/PhysRevD.78.123510} {\bibfield  {journal} {\bibinfo  {journal} {Phys.
  Rev. D}\ }\textbf {\bibinfo {volume} {78}},\ \bibinfo {pages} {123510}
  (\bibinfo {year} {2008})},\ \Eprint {http://arxiv.org/abs/0806.2681}
  {arXiv:0806.2681 [astro-ph]} \BibitemShut {NoStop}%
\bibitem [{\citenamefont {Ilie}\ \emph
  {et~al.}(2020{\natexlab{a}})\citenamefont {Ilie}, \citenamefont {Levy},
  \citenamefont {Pilawa},\ and\ \citenamefont {Zhang}}]{Ilie:2020iup}%
  \BibitemOpen
  \bibfield  {author} {\bibinfo {author} {\bibfnamefont {C.}~\bibnamefont
  {Ilie}}, \bibinfo {author} {\bibfnamefont {C.}~\bibnamefont {Levy}}, \bibinfo
  {author} {\bibfnamefont {J.}~\bibnamefont {Pilawa}}, \ and\ \bibinfo {author}
  {\bibfnamefont {S.}~\bibnamefont {Zhang}},\ }\href@noop {} {\  (\bibinfo
  {year} {2020}{\natexlab{a}})},\ \Eprint {http://arxiv.org/abs/2009.11478}
  {arXiv:2009.11478 [astro-ph.CO]} \BibitemShut {NoStop}%
\bibitem [{\citenamefont {Ilie}\ \emph
  {et~al.}(2020{\natexlab{b}})\citenamefont {Ilie}, \citenamefont {Levy},
  \citenamefont {Pilawa},\ and\ \citenamefont {Zhang}}]{Ilie:2020nzp}%
  \BibitemOpen
  \bibfield  {author} {\bibinfo {author} {\bibfnamefont {C.}~\bibnamefont
  {Ilie}}, \bibinfo {author} {\bibfnamefont {C.}~\bibnamefont {Levy}}, \bibinfo
  {author} {\bibfnamefont {J.}~\bibnamefont {Pilawa}}, \ and\ \bibinfo {author}
  {\bibfnamefont {S.}~\bibnamefont {Zhang}},\ }\href@noop {} {\  (\bibinfo
  {year} {2020}{\natexlab{b}})},\ \Eprint {http://arxiv.org/abs/2009.11474}
  {arXiv:2009.11474 [astro-ph.CO]} \BibitemShut {NoStop}%
\bibitem [{\citenamefont {Leane}\ \emph {et~al.}(2017)\citenamefont {Leane},
  \citenamefont {Ng},\ and\ \citenamefont {Beacom}}]{Leane:2017vag}%
  \BibitemOpen
  \bibfield  {author} {\bibinfo {author} {\bibfnamefont {R.~K.}\ \bibnamefont
  {Leane}}, \bibinfo {author} {\bibfnamefont {K.~C.~Y.}\ \bibnamefont {Ng}}, \
  and\ \bibinfo {author} {\bibfnamefont {J.~F.}\ \bibnamefont {Beacom}},\
  }\href {\doibase 10.1103/PhysRevD.95.123016} {\bibfield  {journal} {\bibinfo
  {journal} {Phys. Rev.}\ }\textbf {\bibinfo {volume} {D95}},\ \bibinfo {pages}
  {123016} (\bibinfo {year} {2017})},\ \Eprint
  {http://arxiv.org/abs/1703.04629} {arXiv:1703.04629 [astro-ph.HE]}
  \BibitemShut {NoStop}%
\bibitem [{\citenamefont {Albert}\ \emph {et~al.}(2018)\citenamefont {Albert}
  \emph {et~al.}}]{HAWC:2018szf}%
  \BibitemOpen
  \bibfield  {author} {\bibinfo {author} {\bibfnamefont {A.}~\bibnamefont
  {Albert}} \emph {et~al.} (\bibinfo {collaboration} {HAWC}),\ }\href {\doibase
  10.1103/PhysRevD.98.123012} {\bibfield  {journal} {\bibinfo  {journal} {Phys.
  Rev. D}\ }\textbf {\bibinfo {volume} {98}},\ \bibinfo {pages} {123012}
  (\bibinfo {year} {2018})},\ \Eprint {http://arxiv.org/abs/1808.05624}
  {arXiv:1808.05624 [hep-ph]} \BibitemShut {NoStop}%
\bibitem [{\citenamefont {Nisa}\ \emph {et~al.}(2019)\citenamefont {Nisa},
  \citenamefont {Beacom}, \citenamefont {BenZvi}, \citenamefont {Leane},
  \citenamefont {Linden}, \citenamefont {Ng}, \citenamefont {Peter},\ and\
  \citenamefont {Zhou}}]{Nisa:2019mpb}%
  \BibitemOpen
  \bibfield  {author} {\bibinfo {author} {\bibfnamefont {M.~U.}\ \bibnamefont
  {Nisa}}, \bibinfo {author} {\bibfnamefont {J.~F.}\ \bibnamefont {Beacom}},
  \bibinfo {author} {\bibfnamefont {S.~Y.}\ \bibnamefont {BenZvi}}, \bibinfo
  {author} {\bibfnamefont {R.~K.}\ \bibnamefont {Leane}}, \bibinfo {author}
  {\bibfnamefont {T.}~\bibnamefont {Linden}}, \bibinfo {author} {\bibfnamefont
  {K.~C.~Y.}\ \bibnamefont {Ng}}, \bibinfo {author} {\bibfnamefont {A.~H.~G.}\
  \bibnamefont {Peter}}, \ and\ \bibinfo {author} {\bibfnamefont
  {B.}~\bibnamefont {Zhou}},\ }\href@noop {} {\  (\bibinfo {year} {2019})},\
  \Eprint {http://arxiv.org/abs/1903.06349} {arXiv:1903.06349 [astro-ph.HE]}
  \BibitemShut {NoStop}%
\bibitem [{\citenamefont {Bell}\ and\ \citenamefont
  {Petraki}(2011)}]{Bell:2011sn}%
  \BibitemOpen
  \bibfield  {author} {\bibinfo {author} {\bibfnamefont {N.~F.}\ \bibnamefont
  {Bell}}\ and\ \bibinfo {author} {\bibfnamefont {K.}~\bibnamefont {Petraki}},\
  }\href {\doibase 10.1088/1475-7516/2011/04/003} {\bibfield  {journal}
  {\bibinfo  {journal} {JCAP}\ }\textbf {\bibinfo {volume} {04}},\ \bibinfo
  {pages} {003} (\bibinfo {year} {2011})},\ \Eprint
  {http://arxiv.org/abs/1102.2958} {arXiv:1102.2958 [hep-ph]} \BibitemShut
  {NoStop}%
\bibitem [{\citenamefont {Choi}\ \emph {et~al.}(2015)\citenamefont {Choi} \emph
  {et~al.}}]{Super-Kamiokande:2015xms}%
  \BibitemOpen
  \bibfield  {author} {\bibinfo {author} {\bibfnamefont {K.}~\bibnamefont
  {Choi}} \emph {et~al.} (\bibinfo {collaboration} {Super-Kamiokande}),\ }\href
  {\doibase 10.1103/PhysRevLett.114.141301} {\bibfield  {journal} {\bibinfo
  {journal} {Phys. Rev. Lett.}\ }\textbf {\bibinfo {volume} {114}},\ \bibinfo
  {pages} {141301} (\bibinfo {year} {2015})},\ \Eprint
  {http://arxiv.org/abs/1503.04858} {arXiv:1503.04858 [hep-ex]} \BibitemShut
  {NoStop}%
\bibitem [{\citenamefont {Aartsen}\ \emph {et~al.}(2017)\citenamefont {Aartsen}
  \emph {et~al.}}]{IceCube:2016dgk}%
  \BibitemOpen
  \bibfield  {author} {\bibinfo {author} {\bibfnamefont {M.~G.}\ \bibnamefont
  {Aartsen}} \emph {et~al.} (\bibinfo {collaboration} {IceCube}),\ }\href
  {\doibase 10.1140/epjc/s10052-017-4689-9} {\bibfield  {journal} {\bibinfo
  {journal} {Eur. Phys. J. C}\ }\textbf {\bibinfo {volume} {77}},\ \bibinfo
  {pages} {146} (\bibinfo {year} {2017})},\ \bibinfo {note} {[Erratum:
  Eur.Phys.J.C 79, 214 (2019)]},\ \Eprint {http://arxiv.org/abs/1612.05949}
  {arXiv:1612.05949 [astro-ph.HE]} \BibitemShut {NoStop}%
\bibitem [{\citenamefont {Adrian-Martinez}\ \emph {et~al.}(2016)\citenamefont
  {Adrian-Martinez} \emph {et~al.}}]{ANTARES:2016xuh}%
  \BibitemOpen
  \bibfield  {author} {\bibinfo {author} {\bibfnamefont {S.}~\bibnamefont
  {Adrian-Martinez}} \emph {et~al.} (\bibinfo {collaboration} {ANTARES}),\
  }\href {\doibase 10.1016/j.physletb.2016.05.019} {\bibfield  {journal}
  {\bibinfo  {journal} {Phys. Lett. B}\ }\textbf {\bibinfo {volume} {759}},\
  \bibinfo {pages} {69} (\bibinfo {year} {2016})},\ \Eprint
  {http://arxiv.org/abs/1603.02228} {arXiv:1603.02228 [astro-ph.HE]}
  \BibitemShut {NoStop}%
\bibitem [{\citenamefont {Leane}\ \emph {et~al.}(2021)\citenamefont {Leane},
  \citenamefont {Linden}, \citenamefont {Mukhopadhyay},\ and\ \citenamefont
  {Toro}}]{Leane:2021ihh}%
  \BibitemOpen
  \bibfield  {author} {\bibinfo {author} {\bibfnamefont {R.~K.}\ \bibnamefont
  {Leane}}, \bibinfo {author} {\bibfnamefont {T.}~\bibnamefont {Linden}},
  \bibinfo {author} {\bibfnamefont {P.}~\bibnamefont {Mukhopadhyay}}, \ and\
  \bibinfo {author} {\bibfnamefont {N.}~\bibnamefont {Toro}},\ }\href {\doibase
  10.1103/PhysRevD.103.075030} {\bibfield  {journal} {\bibinfo  {journal}
  {Phys. Rev. D}\ }\textbf {\bibinfo {volume} {103}},\ \bibinfo {pages}
  {075030} (\bibinfo {year} {2021})},\ \Eprint
  {http://arxiv.org/abs/2101.12213} {arXiv:2101.12213 [astro-ph.HE]}
  \BibitemShut {NoStop}%
\bibitem [{\citenamefont {Leane}\ and\ \citenamefont
  {Linden}(2021)}]{Leane:2021tjj}%
  \BibitemOpen
  \bibfield  {author} {\bibinfo {author} {\bibfnamefont {R.~K.}\ \bibnamefont
  {Leane}}\ and\ \bibinfo {author} {\bibfnamefont {T.}~\bibnamefont {Linden}},\
  }\href@noop {} {\  (\bibinfo {year} {2021})},\ \Eprint
  {http://arxiv.org/abs/2104.02068} {arXiv:2104.02068 [astro-ph.HE]}
  \BibitemShut {NoStop}%
\bibitem [{\citenamefont {Bose}\ \emph {et~al.}(2021)\citenamefont {Bose},
  \citenamefont {Maity},\ and\ \citenamefont {Ray}}]{Bose:2021yhz}%
  \BibitemOpen
  \bibfield  {author} {\bibinfo {author} {\bibfnamefont {D.}~\bibnamefont
  {Bose}}, \bibinfo {author} {\bibfnamefont {T.~N.}\ \bibnamefont {Maity}}, \
  and\ \bibinfo {author} {\bibfnamefont {T.~S.}\ \bibnamefont {Ray}},\
  }\href@noop {} {\  (\bibinfo {year} {2021})},\ \Eprint
  {http://arxiv.org/abs/2108.12420} {arXiv:2108.12420 [hep-ph]} \BibitemShut
  {NoStop}%
\bibitem [{\citenamefont {Goldman}\ and\ \citenamefont
  {Nussinov}(1989{\natexlab{b}})}]{PhysRevD.40.3221}%
  \BibitemOpen
  \bibfield  {author} {\bibinfo {author} {\bibfnamefont {I.}~\bibnamefont
  {Goldman}}\ and\ \bibinfo {author} {\bibfnamefont {S.}~\bibnamefont
  {Nussinov}},\ }\href {\doibase 10.1103/PhysRevD.40.3221} {\bibfield
  {journal} {\bibinfo  {journal} {Phys. Rev. D}\ }\textbf {\bibinfo {volume}
  {40}},\ \bibinfo {pages} {3221} (\bibinfo {year}
  {1989}{\natexlab{b}})}\BibitemShut {NoStop}%
\bibitem [{\citenamefont {Kouvaris}(2008)}]{Kouvaris:2007ay}%
  \BibitemOpen
  \bibfield  {author} {\bibinfo {author} {\bibfnamefont {C.}~\bibnamefont
  {Kouvaris}},\ }\href {\doibase 10.1103/PhysRevD.77.023006} {\bibfield
  {journal} {\bibinfo  {journal} {Phys. Rev. D}\ }\textbf {\bibinfo {volume}
  {77}},\ \bibinfo {pages} {023006} (\bibinfo {year} {2008})},\ \Eprint
  {http://arxiv.org/abs/0708.2362} {arXiv:0708.2362 [astro-ph]} \BibitemShut
  {NoStop}%
\bibitem [{\citenamefont {Kouvaris}\ and\ \citenamefont
  {Tinyakov}(2010{\natexlab{b}})}]{Kouvaris_2010}%
  \BibitemOpen
  \bibfield  {author} {\bibinfo {author} {\bibfnamefont {C.}~\bibnamefont
  {Kouvaris}}\ and\ \bibinfo {author} {\bibfnamefont {P.}~\bibnamefont
  {Tinyakov}},\ }\href {\doibase 10.1103/physrevd.82.063531} {\bibfield
  {journal} {\bibinfo  {journal} {Physical Review D}\ }\textbf {\bibinfo
  {volume} {82}} (\bibinfo {year} {2010}{\natexlab{b}}),\
  10.1103/physrevd.82.063531}\BibitemShut {NoStop}%
\bibitem [{\citenamefont {Kouvaris}\ and\ \citenamefont
  {Tinyakov}(2011)}]{Kouvaris:2011fi}%
  \BibitemOpen
  \bibfield  {author} {\bibinfo {author} {\bibfnamefont {C.}~\bibnamefont
  {Kouvaris}}\ and\ \bibinfo {author} {\bibfnamefont {P.}~\bibnamefont
  {Tinyakov}},\ }\href {\doibase 10.1103/PhysRevLett.107.091301} {\bibfield
  {journal} {\bibinfo  {journal} {Phys. Rev. Lett.}\ }\textbf {\bibinfo
  {volume} {107}},\ \bibinfo {pages} {091301} (\bibinfo {year} {2011})},\
  \Eprint {http://arxiv.org/abs/1104.0382} {arXiv:1104.0382 [astro-ph.CO]}
  \BibitemShut {NoStop}%
\bibitem [{\citenamefont {McDermott}\ \emph {et~al.}(2012)\citenamefont
  {McDermott}, \citenamefont {Yu},\ and\ \citenamefont
  {Zurek}}]{McDermott:2011jp}%
  \BibitemOpen
  \bibfield  {author} {\bibinfo {author} {\bibfnamefont {S.~D.}\ \bibnamefont
  {McDermott}}, \bibinfo {author} {\bibfnamefont {H.-B.}\ \bibnamefont {Yu}}, \
  and\ \bibinfo {author} {\bibfnamefont {K.~M.}\ \bibnamefont {Zurek}},\ }\href
  {\doibase 10.1103/PhysRevD.85.023519} {\bibfield  {journal} {\bibinfo
  {journal} {Phys. Rev.}\ }\textbf {\bibinfo {volume} {D85}},\ \bibinfo {pages}
  {023519} (\bibinfo {year} {2012})},\ \Eprint {http://arxiv.org/abs/1103.5472}
  {arXiv:1103.5472 [hep-ph]} \BibitemShut {NoStop}%
\bibitem [{\citenamefont {Guver}\ \emph {et~al.}(2014)\citenamefont {Guver},
  \citenamefont {Erkoca}, \citenamefont {Hall~Reno},\ and\ \citenamefont
  {Sarcevic}}]{Guver:2012ba}%
  \BibitemOpen
  \bibfield  {author} {\bibinfo {author} {\bibfnamefont {T.}~\bibnamefont
  {Guver}}, \bibinfo {author} {\bibfnamefont {A.~E.}\ \bibnamefont {Erkoca}},
  \bibinfo {author} {\bibfnamefont {M.}~\bibnamefont {Hall~Reno}}, \ and\
  \bibinfo {author} {\bibfnamefont {I.}~\bibnamefont {Sarcevic}},\ }\href
  {\doibase 10.1088/1475-7516/2014/05/013} {\bibfield  {journal} {\bibinfo
  {journal} {JCAP}\ }\textbf {\bibinfo {volume} {1405}},\ \bibinfo {pages}
  {013} (\bibinfo {year} {2014})},\ \Eprint {http://arxiv.org/abs/1201.2400}
  {arXiv:1201.2400 [hep-ph]} \BibitemShut {NoStop}%
\bibitem [{\citenamefont {Kouvaris}\ and\ \citenamefont
  {Tinyakov}(2013)}]{Kouvaris:2012dz}%
  \BibitemOpen
  \bibfield  {author} {\bibinfo {author} {\bibfnamefont {C.}~\bibnamefont
  {Kouvaris}}\ and\ \bibinfo {author} {\bibfnamefont {P.}~\bibnamefont
  {Tinyakov}},\ }\href {\doibase 10.1103/PhysRevD.87.123537} {\bibfield
  {journal} {\bibinfo  {journal} {Phys. Rev. D}\ }\textbf {\bibinfo {volume}
  {87}},\ \bibinfo {pages} {123537} (\bibinfo {year} {2013})},\ \Eprint
  {http://arxiv.org/abs/1212.4075} {arXiv:1212.4075 [astro-ph.HE]} \BibitemShut
  {NoStop}%
\bibitem [{\citenamefont {Bramante}\ \emph {et~al.}(2013)\citenamefont
  {Bramante}, \citenamefont {Fukushima},\ and\ \citenamefont
  {Kumar}}]{Bramante:2013hn}%
  \BibitemOpen
  \bibfield  {author} {\bibinfo {author} {\bibfnamefont {J.}~\bibnamefont
  {Bramante}}, \bibinfo {author} {\bibfnamefont {K.}~\bibnamefont {Fukushima}},
  \ and\ \bibinfo {author} {\bibfnamefont {J.}~\bibnamefont {Kumar}},\ }\href
  {\doibase 10.1103/PhysRevD.87.055012} {\bibfield  {journal} {\bibinfo
  {journal} {Phys. Rev.}\ }\textbf {\bibinfo {volume} {D87}},\ \bibinfo {pages}
  {055012} (\bibinfo {year} {2013})},\ \Eprint {http://arxiv.org/abs/1301.0036}
  {arXiv:1301.0036 [hep-ph]} \BibitemShut {NoStop}%
\bibitem [{\citenamefont {Bell}\ \emph {et~al.}(2013)\citenamefont {Bell},
  \citenamefont {Melatos},\ and\ \citenamefont {Petraki}}]{Bell:2013xk}%
  \BibitemOpen
  \bibfield  {author} {\bibinfo {author} {\bibfnamefont {N.~F.}\ \bibnamefont
  {Bell}}, \bibinfo {author} {\bibfnamefont {A.}~\bibnamefont {Melatos}}, \
  and\ \bibinfo {author} {\bibfnamefont {K.}~\bibnamefont {Petraki}},\ }\href
  {\doibase 10.1103/PhysRevD.87.123507} {\bibfield  {journal} {\bibinfo
  {journal} {Phys. Rev.}\ }\textbf {\bibinfo {volume} {D87}},\ \bibinfo {pages}
  {123507} (\bibinfo {year} {2013})},\ \Eprint {http://arxiv.org/abs/1301.6811}
  {arXiv:1301.6811 [hep-ph]} \BibitemShut {NoStop}%
\bibitem [{\citenamefont {Bramante}\ \emph {et~al.}(2014)\citenamefont
  {Bramante}, \citenamefont {Fukushima}, \citenamefont {Kumar},\ and\
  \citenamefont {Stopnitzky}}]{Bramante:2013nma}%
  \BibitemOpen
  \bibfield  {author} {\bibinfo {author} {\bibfnamefont {J.}~\bibnamefont
  {Bramante}}, \bibinfo {author} {\bibfnamefont {K.}~\bibnamefont {Fukushima}},
  \bibinfo {author} {\bibfnamefont {J.}~\bibnamefont {Kumar}}, \ and\ \bibinfo
  {author} {\bibfnamefont {E.}~\bibnamefont {Stopnitzky}},\ }\href {\doibase
  10.1103/PhysRevD.89.015010} {\bibfield  {journal} {\bibinfo  {journal} {Phys.
  Rev.}\ }\textbf {\bibinfo {volume} {D89}},\ \bibinfo {pages} {015010}
  (\bibinfo {year} {2014})},\ \Eprint {http://arxiv.org/abs/1310.3509}
  {arXiv:1310.3509 [hep-ph]} \BibitemShut {NoStop}%
\bibitem [{\citenamefont {Bramante}\ and\ \citenamefont
  {Linden}(2014)}]{Bramante:2014zca}%
  \BibitemOpen
  \bibfield  {author} {\bibinfo {author} {\bibfnamefont {J.}~\bibnamefont
  {Bramante}}\ and\ \bibinfo {author} {\bibfnamefont {T.}~\bibnamefont
  {Linden}},\ }\href {\doibase 10.1103/PhysRevLett.113.191301} {\bibfield
  {journal} {\bibinfo  {journal} {Phys. Rev. Lett.}\ }\textbf {\bibinfo
  {volume} {113}},\ \bibinfo {pages} {191301} (\bibinfo {year} {2014})},\
  \Eprint {http://arxiv.org/abs/1405.1031} {arXiv:1405.1031 [astro-ph.HE]}
  \BibitemShut {NoStop}%
\bibitem [{\citenamefont {Bramante}\ and\ \citenamefont
  {Elahi}(2015)}]{Bramante:2015dfa}%
  \BibitemOpen
  \bibfield  {author} {\bibinfo {author} {\bibfnamefont {J.}~\bibnamefont
  {Bramante}}\ and\ \bibinfo {author} {\bibfnamefont {F.}~\bibnamefont
  {Elahi}},\ }\href {\doibase 10.1103/PhysRevD.91.115001} {\bibfield  {journal}
  {\bibinfo  {journal} {Phys. Rev.}\ }\textbf {\bibinfo {volume} {D91}},\
  \bibinfo {pages} {115001} (\bibinfo {year} {2015})},\ \Eprint
  {http://arxiv.org/abs/1504.04019} {arXiv:1504.04019 [hep-ph]} \BibitemShut
  {NoStop}%
\bibitem [{\citenamefont {zhong Fan}\ \emph {et~al.}(2012)\citenamefont {zhong
  Fan}, \citenamefont {zhi Yang},\ and\ \citenamefont
  {Chang}}]{fan2012constraining}%
  \BibitemOpen
  \bibfield  {author} {\bibinfo {author} {\bibfnamefont {Y.}~\bibnamefont
  {zhong Fan}}, \bibinfo {author} {\bibfnamefont {R.}~\bibnamefont {zhi Yang}},
  \ and\ \bibinfo {author} {\bibfnamefont {J.}~\bibnamefont {Chang}},\
  }\href@noop {} {} (\bibinfo {year} {2012}),\ \Eprint
  {http://arxiv.org/abs/1204.2564} {arXiv:1204.2564 [astro-ph.HE]} \BibitemShut
  {NoStop}%
\bibitem [{\citenamefont {Ellis}\ \emph
  {et~al.}(2018{\natexlab{a}})\citenamefont {Ellis}, \citenamefont {Hektor},
  \citenamefont {H\"utsi}, \citenamefont {Kannike}, \citenamefont {Marzola},
  \citenamefont {Raidal},\ and\ \citenamefont {Vaskonen}}]{Ellis:2017jgp}%
  \BibitemOpen
  \bibfield  {author} {\bibinfo {author} {\bibfnamefont {J.}~\bibnamefont
  {Ellis}}, \bibinfo {author} {\bibfnamefont {A.}~\bibnamefont {Hektor}},
  \bibinfo {author} {\bibfnamefont {G.}~\bibnamefont {H\"utsi}}, \bibinfo
  {author} {\bibfnamefont {K.}~\bibnamefont {Kannike}}, \bibinfo {author}
  {\bibfnamefont {L.}~\bibnamefont {Marzola}}, \bibinfo {author} {\bibfnamefont
  {M.}~\bibnamefont {Raidal}}, \ and\ \bibinfo {author} {\bibfnamefont
  {V.}~\bibnamefont {Vaskonen}},\ }\href {\doibase
  10.1016/j.physletb.2018.04.048} {\bibfield  {journal} {\bibinfo  {journal}
  {Phys. Lett. B}\ }\textbf {\bibinfo {volume} {781}},\ \bibinfo {pages} {607}
  (\bibinfo {year} {2018}{\natexlab{a}})},\ \Eprint
  {http://arxiv.org/abs/1710.05540} {arXiv:1710.05540 [astro-ph.CO]}
  \BibitemShut {NoStop}%
\bibitem [{\citenamefont {Croon}\ \emph {et~al.}(2018)\citenamefont {Croon},
  \citenamefont {Nelson}, \citenamefont {Sun}, \citenamefont {Walker},\ and\
  \citenamefont {Xianyu}}]{Croon:2017zcu}%
  \BibitemOpen
  \bibfield  {author} {\bibinfo {author} {\bibfnamefont {D.}~\bibnamefont
  {Croon}}, \bibinfo {author} {\bibfnamefont {A.~E.}\ \bibnamefont {Nelson}},
  \bibinfo {author} {\bibfnamefont {C.}~\bibnamefont {Sun}}, \bibinfo {author}
  {\bibfnamefont {D.~G.~E.}\ \bibnamefont {Walker}}, \ and\ \bibinfo {author}
  {\bibfnamefont {Z.-Z.}\ \bibnamefont {Xianyu}},\ }\href {\doibase
  10.3847/2041-8213/aabe76} {\bibfield  {journal} {\bibinfo  {journal}
  {Astrophys. J. Lett.}\ }\textbf {\bibinfo {volume} {858}},\ \bibinfo {pages}
  {L2} (\bibinfo {year} {2018})},\ \Eprint {http://arxiv.org/abs/1711.02096}
  {arXiv:1711.02096 [hep-ph]} \BibitemShut {NoStop}%
\bibitem [{\citenamefont {Li}\ \emph {et~al.}(2012{\natexlab{a}})\citenamefont
  {Li}, \citenamefont {Huang},\ and\ \citenamefont {Xu}}]{Li:2012ii}%
  \BibitemOpen
  \bibfield  {author} {\bibinfo {author} {\bibfnamefont {A.}~\bibnamefont
  {Li}}, \bibinfo {author} {\bibfnamefont {F.}~\bibnamefont {Huang}}, \ and\
  \bibinfo {author} {\bibfnamefont {R.-X.}\ \bibnamefont {Xu}},\ }\href
  {\doibase 10.1016/j.astropartphys.2012.07.006} {\bibfield  {journal}
  {\bibinfo  {journal} {Astropart. Phys.}\ }\textbf {\bibinfo {volume} {37}},\
  \bibinfo {pages} {70} (\bibinfo {year} {2012}{\natexlab{a}})},\ \Eprint
  {http://arxiv.org/abs/1208.3722} {arXiv:1208.3722 [astro-ph.SR]} \BibitemShut
  {NoStop}%
\bibitem [{\citenamefont {Leung}\ \emph {et~al.}(2011)\citenamefont {Leung},
  \citenamefont {Chu},\ and\ \citenamefont {Lin}}]{Leung:2011zz}%
  \BibitemOpen
  \bibfield  {author} {\bibinfo {author} {\bibfnamefont {S.~C.}\ \bibnamefont
  {Leung}}, \bibinfo {author} {\bibfnamefont {M.~C.}\ \bibnamefont {Chu}}, \
  and\ \bibinfo {author} {\bibfnamefont {L.~M.}\ \bibnamefont {Lin}},\ }\href
  {\doibase 10.1103/PhysRevD.84.107301} {\bibfield  {journal} {\bibinfo
  {journal} {Phys. Rev. D}\ }\textbf {\bibinfo {volume} {84}},\ \bibinfo
  {pages} {107301} (\bibinfo {year} {2011})},\ \Eprint
  {http://arxiv.org/abs/1111.1787} {arXiv:1111.1787 [astro-ph.CO]} \BibitemShut
  {NoStop}%
\bibitem [{\citenamefont {Xiang}\ \emph {et~al.}(2014)\citenamefont {Xiang},
  \citenamefont {Jiang}, \citenamefont {Zhang},\ and\ \citenamefont
  {Yang}}]{Xiang:2013xwa}%
  \BibitemOpen
  \bibfield  {author} {\bibinfo {author} {\bibfnamefont {Q.-F.}\ \bibnamefont
  {Xiang}}, \bibinfo {author} {\bibfnamefont {W.-Z.}\ \bibnamefont {Jiang}},
  \bibinfo {author} {\bibfnamefont {D.-R.}\ \bibnamefont {Zhang}}, \ and\
  \bibinfo {author} {\bibfnamefont {R.-Y.}\ \bibnamefont {Yang}},\ }\href
  {\doibase 10.1103/PhysRevC.89.025803} {\bibfield  {journal} {\bibinfo
  {journal} {Phys. Rev. C}\ }\textbf {\bibinfo {volume} {89}},\ \bibinfo
  {pages} {025803} (\bibinfo {year} {2014})},\ \Eprint
  {http://arxiv.org/abs/1305.7354} {arXiv:1305.7354 [astro-ph.SR]} \BibitemShut
  {NoStop}%
\bibitem [{\citenamefont {Tolos}\ \emph {et~al.}(2015)\citenamefont {Tolos},
  \citenamefont {Schaffner-Bielich},\ and\ \citenamefont
  {Dengler}}]{Tolos:2015qra}%
  \BibitemOpen
  \bibfield  {author} {\bibinfo {author} {\bibfnamefont {L.}~\bibnamefont
  {Tolos}}, \bibinfo {author} {\bibfnamefont {J.}~\bibnamefont
  {Schaffner-Bielich}}, \ and\ \bibinfo {author} {\bibfnamefont
  {Y.}~\bibnamefont {Dengler}},\ }\href {\doibase 10.1103/PhysRevD.92.123002}
  {\bibfield  {journal} {\bibinfo  {journal} {Phys. Rev. D}\ }\textbf {\bibinfo
  {volume} {92}},\ \bibinfo {pages} {123002} (\bibinfo {year} {2015})},\
  \bibinfo {note} {[Erratum: Phys.Rev.D 103, 109901 (2021)]},\ \Eprint
  {http://arxiv.org/abs/1507.08197} {arXiv:1507.08197 [astro-ph.HE]}
  \BibitemShut {NoStop}%
\bibitem [{\citenamefont {Panotopoulos}\ and\ \citenamefont
  {Lopes}(2017)}]{Panotopoulos:2017idn}%
  \BibitemOpen
  \bibfield  {author} {\bibinfo {author} {\bibfnamefont {G.}~\bibnamefont
  {Panotopoulos}}\ and\ \bibinfo {author} {\bibfnamefont {I.}~\bibnamefont
  {Lopes}},\ }\href {\doibase 10.1103/PhysRevD.96.083004} {\bibfield  {journal}
  {\bibinfo  {journal} {Phys. Rev. D}\ }\textbf {\bibinfo {volume} {96}},\
  \bibinfo {pages} {083004} (\bibinfo {year} {2017})},\ \Eprint
  {http://arxiv.org/abs/1709.06312} {arXiv:1709.06312 [hep-ph]} \BibitemShut
  {NoStop}%
\bibitem [{\citenamefont {Gresham}\ and\ \citenamefont
  {Zurek}(2019)}]{Gresham:2018rqo}%
  \BibitemOpen
  \bibfield  {author} {\bibinfo {author} {\bibfnamefont {M.~I.}\ \bibnamefont
  {Gresham}}\ and\ \bibinfo {author} {\bibfnamefont {K.~M.}\ \bibnamefont
  {Zurek}},\ }\href {\doibase 10.1103/PhysRevD.99.083008} {\bibfield  {journal}
  {\bibinfo  {journal} {Phys. Rev. D}\ }\textbf {\bibinfo {volume} {99}},\
  \bibinfo {pages} {083008} (\bibinfo {year} {2019})},\ \Eprint
  {http://arxiv.org/abs/1809.08254} {arXiv:1809.08254 [astro-ph.CO]}
  \BibitemShut {NoStop}%
\bibitem [{\citenamefont {Karkevandi}\ \emph {et~al.}(2022)\citenamefont
  {Karkevandi}, \citenamefont {Shakeri}, \citenamefont {Sagun},\ and\
  \citenamefont {Ivanytskyi}}]{Karkevandi:2021ygv}%
  \BibitemOpen
  \bibfield  {author} {\bibinfo {author} {\bibfnamefont {D.~R.}\ \bibnamefont
  {Karkevandi}}, \bibinfo {author} {\bibfnamefont {S.}~\bibnamefont {Shakeri}},
  \bibinfo {author} {\bibfnamefont {V.}~\bibnamefont {Sagun}}, \ and\ \bibinfo
  {author} {\bibfnamefont {O.}~\bibnamefont {Ivanytskyi}},\ }\href {\doibase
  10.1103/PhysRevD.105.023001} {\bibfield  {journal} {\bibinfo  {journal}
  {Phys. Rev. D}\ }\textbf {\bibinfo {volume} {105}},\ \bibinfo {pages}
  {023001} (\bibinfo {year} {2022})},\ \Eprint
  {http://arxiv.org/abs/2109.03801} {arXiv:2109.03801 [astro-ph.HE]}
  \BibitemShut {NoStop}%
\bibitem [{\citenamefont {Nelson}\ \emph {et~al.}(2019)\citenamefont {Nelson},
  \citenamefont {Reddy},\ and\ \citenamefont {Zhou}}]{Nelson:2018xtr}%
  \BibitemOpen
  \bibfield  {author} {\bibinfo {author} {\bibfnamefont {A.}~\bibnamefont
  {Nelson}}, \bibinfo {author} {\bibfnamefont {S.}~\bibnamefont {Reddy}}, \
  and\ \bibinfo {author} {\bibfnamefont {D.}~\bibnamefont {Zhou}},\ }\href
  {\doibase 10.1088/1475-7516/2019/07/012} {\bibfield  {journal} {\bibinfo
  {journal} {JCAP}\ }\textbf {\bibinfo {volume} {1907}},\ \bibinfo {pages}
  {012} (\bibinfo {year} {2019})},\ \Eprint {http://arxiv.org/abs/1803.03266}
  {arXiv:1803.03266 [hep-ph]} \BibitemShut {NoStop}%
\bibitem [{\citenamefont {Dengler}\ \emph {et~al.}(2022)\citenamefont
  {Dengler}, \citenamefont {Schaffner-Bielich},\ and\ \citenamefont
  {Tolos}}]{Dengler:2021qcq}%
  \BibitemOpen
  \bibfield  {author} {\bibinfo {author} {\bibfnamefont {Y.}~\bibnamefont
  {Dengler}}, \bibinfo {author} {\bibfnamefont {J.}~\bibnamefont
  {Schaffner-Bielich}}, \ and\ \bibinfo {author} {\bibfnamefont
  {L.}~\bibnamefont {Tolos}},\ }\href {\doibase 10.1103/PhysRevD.105.043013}
  {\bibfield  {journal} {\bibinfo  {journal} {Phys. Rev. D}\ }\textbf {\bibinfo
  {volume} {105}},\ \bibinfo {pages} {043013} (\bibinfo {year} {2022})},\
  \Eprint {http://arxiv.org/abs/2111.06197} {arXiv:2111.06197 [astro-ph.HE]}
  \BibitemShut {NoStop}%
\bibitem [{\citenamefont {Ellis}\ \emph
  {et~al.}(2018{\natexlab{b}})\citenamefont {Ellis}, \citenamefont {Hütsi},
  \citenamefont {Kannike}, \citenamefont {Marzola}, \citenamefont {Raidal},\
  and\ \citenamefont {Vaskonen}}]{Ellis:2018bkr}%
  \BibitemOpen
  \bibfield  {author} {\bibinfo {author} {\bibfnamefont {J.}~\bibnamefont
  {Ellis}}, \bibinfo {author} {\bibfnamefont {G.}~\bibnamefont {Hütsi}},
  \bibinfo {author} {\bibfnamefont {K.}~\bibnamefont {Kannike}}, \bibinfo
  {author} {\bibfnamefont {L.}~\bibnamefont {Marzola}}, \bibinfo {author}
  {\bibfnamefont {M.}~\bibnamefont {Raidal}}, \ and\ \bibinfo {author}
  {\bibfnamefont {V.}~\bibnamefont {Vaskonen}},\ }\href {\doibase
  10.1103/PhysRevD.97.123007} {\bibfield  {journal} {\bibinfo  {journal}
  {Phys.\ Rev.\ D}\ }\textbf {\bibinfo {volume} {97}},\ \bibinfo {pages}
  {123007} (\bibinfo {year} {2018}{\natexlab{b}})},\ \Eprint
  {http://arxiv.org/abs/1804.01418} {arXiv:1804.01418 [astro-ph.CO]}
  \BibitemShut {NoStop}%
\bibitem [{\citenamefont {Ciarcelluti}\ and\ \citenamefont
  {Sandin}(2011)}]{Ciarcelluti:2010ji}%
  \BibitemOpen
  \bibfield  {author} {\bibinfo {author} {\bibfnamefont {P.}~\bibnamefont
  {Ciarcelluti}}\ and\ \bibinfo {author} {\bibfnamefont {F.}~\bibnamefont
  {Sandin}},\ }\href {\doibase 10.1016/j.physletb.2010.11.021} {\bibfield
  {journal} {\bibinfo  {journal} {Phys. Lett. B}\ }\textbf {\bibinfo {volume}
  {695}},\ \bibinfo {pages} {19} (\bibinfo {year} {2011})},\ \Eprint
  {http://arxiv.org/abs/1005.0857} {arXiv:1005.0857 [astro-ph.HE]} \BibitemShut
  {NoStop}%
\bibitem [{\citenamefont {Gupta}\ \emph {et~al.}(2022)\citenamefont {Gupta},
  \citenamefont {Puecher}, \citenamefont {Pang}, \citenamefont {Janquart},
  \citenamefont {Koekoek},\ and\ \citenamefont {Broeck
  Van~Den}}]{Gupta:2022qgg}%
  \BibitemOpen
  \bibfield  {author} {\bibinfo {author} {\bibfnamefont {P.~K.}\ \bibnamefont
  {Gupta}}, \bibinfo {author} {\bibfnamefont {A.}~\bibnamefont {Puecher}},
  \bibinfo {author} {\bibfnamefont {P.~T.~H.}\ \bibnamefont {Pang}}, \bibinfo
  {author} {\bibfnamefont {J.}~\bibnamefont {Janquart}}, \bibinfo {author}
  {\bibfnamefont {G.}~\bibnamefont {Koekoek}}, \ and\ \bibinfo {author}
  {\bibfnamefont {C.}~\bibnamefont {Broeck Van~Den}},\ }\href@noop {} {\
  (\bibinfo {year} {2022})},\ \Eprint {http://arxiv.org/abs/2205.01182}
  {arXiv:2205.01182 [gr-qc]} \BibitemShut {NoStop}%
\bibitem [{\citenamefont {Arun}\ \emph {et~al.}(2022)\citenamefont {Arun} \emph
  {et~al.}}]{LISA:2022kgy}%
  \BibitemOpen
  \bibfield  {author} {\bibinfo {author} {\bibfnamefont {K.~G.}\ \bibnamefont
  {Arun}} \emph {et~al.} (\bibinfo {collaboration} {LISA}),\ }\href@noop {} {\
  (\bibinfo {year} {2022})},\ \Eprint {http://arxiv.org/abs/2205.01597}
  {arXiv:2205.01597 [gr-qc]} \BibitemShut {NoStop}%
\bibitem [{\citenamefont {Oppenheimer}\ and\ \citenamefont
  {Volkoff}(1939)}]{Oppenheimer:1939ne}%
  \BibitemOpen
  \bibfield  {author} {\bibinfo {author} {\bibfnamefont {J.~R.}\ \bibnamefont
  {Oppenheimer}}\ and\ \bibinfo {author} {\bibfnamefont {G.~M.}\ \bibnamefont
  {Volkoff}},\ }\href {\doibase 10.1103/PhysRev.55.374} {\bibfield  {journal}
  {\bibinfo  {journal} {Phys. Rev.}\ }\textbf {\bibinfo {volume} {55}},\
  \bibinfo {pages} {374} (\bibinfo {year} {1939})}\BibitemShut {NoStop}%
\bibitem [{\citenamefont {Tolman}(1934)}]{Tolman:1934za}%
  \BibitemOpen
  \bibfield  {author} {\bibinfo {author} {\bibfnamefont {R.~C.}\ \bibnamefont
  {Tolman}},\ }\href {\doibase 10.1073/pnas.20.3.169} {\bibfield  {journal}
  {\bibinfo  {journal} {Proc. Nat. Acad. Sci.}\ }\textbf {\bibinfo {volume}
  {20}},\ \bibinfo {pages} {169} (\bibinfo {year} {1934})}\BibitemShut
  {NoStop}%
\bibitem [{\citenamefont {Pearson}\ \emph {et~al.}(2018)\citenamefont
  {Pearson}, \citenamefont {Chamel}, \citenamefont {Potekhin}, \citenamefont
  {Fantina}, \citenamefont {Ducoin}, \citenamefont {Dutta},\ and\ \citenamefont
  {Goriely}}]{Pearson:2018tkr}%
  \BibitemOpen
  \bibfield  {author} {\bibinfo {author} {\bibfnamefont {J.~M.}\ \bibnamefont
  {Pearson}}, \bibinfo {author} {\bibfnamefont {N.}~\bibnamefont {Chamel}},
  \bibinfo {author} {\bibfnamefont {A.~Y.}\ \bibnamefont {Potekhin}}, \bibinfo
  {author} {\bibfnamefont {A.~F.}\ \bibnamefont {Fantina}}, \bibinfo {author}
  {\bibfnamefont {C.}~\bibnamefont {Ducoin}}, \bibinfo {author} {\bibfnamefont
  {A.~K.}\ \bibnamefont {Dutta}}, \ and\ \bibinfo {author} {\bibfnamefont
  {S.}~\bibnamefont {Goriely}},\ }\href {\doibase 10.1093/mnras/sty2413,
  10.1093/mnras/stz800} {\bibfield  {journal} {\bibinfo  {journal} {Mon. Not.
  Roy. Astron. Soc.}\ }\textbf {\bibinfo {volume} {481}},\ \bibinfo {pages}
  {2994} (\bibinfo {year} {2018})},\ \bibinfo {note} {[erratum: Mon. Not. Roy.
  Astron. Soc.486,no.1,768(2019)]},\ \Eprint {http://arxiv.org/abs/1903.04981}
  {arXiv:1903.04981 [astro-ph.HE]} \BibitemShut {NoStop}%
\bibitem [{\citenamefont {Goriely}\ \emph {et~al.}(2013)\citenamefont
  {Goriely}, \citenamefont {Chamel},\ and\ \citenamefont
  {Pearson}}]{Goriely:2013xba}%
  \BibitemOpen
  \bibfield  {author} {\bibinfo {author} {\bibfnamefont {S.}~\bibnamefont
  {Goriely}}, \bibinfo {author} {\bibfnamefont {N.}~\bibnamefont {Chamel}}, \
  and\ \bibinfo {author} {\bibfnamefont {J.~M.}\ \bibnamefont {Pearson}},\
  }\href {\doibase 10.1103/PhysRevC.88.024308} {\bibfield  {journal} {\bibinfo
  {journal} {Phys. Rev.}\ }\textbf {\bibinfo {volume} {C88}},\ \bibinfo {pages}
  {024308} (\bibinfo {year} {2013})}\BibitemShut {NoStop}%
\bibitem [{\citenamefont {Audi}\ \emph {et~al.}(2012)\citenamefont {Audi},
  \citenamefont {Wang}, \citenamefont {Wapstra}, \citenamefont {Kondev},
  \citenamefont {MacCormick}, \citenamefont {Xu},\ and\ \citenamefont
  {Pfeiffer}}]{audi2012ame2012}%
  \BibitemOpen
  \bibfield  {author} {\bibinfo {author} {\bibfnamefont {G.}~\bibnamefont
  {Audi}}, \bibinfo {author} {\bibfnamefont {M.}~\bibnamefont {Wang}}, \bibinfo
  {author} {\bibfnamefont {A.}~\bibnamefont {Wapstra}}, \bibinfo {author}
  {\bibfnamefont {F.}~\bibnamefont {Kondev}}, \bibinfo {author} {\bibfnamefont
  {M.}~\bibnamefont {MacCormick}}, \bibinfo {author} {\bibfnamefont
  {X.}~\bibnamefont {Xu}}, \ and\ \bibinfo {author} {\bibfnamefont
  {B.}~\bibnamefont {Pfeiffer}},\ }\href@noop {} {\bibfield  {journal}
  {\bibinfo  {journal} {Chinese physics C}\ }\textbf {\bibinfo {volume} {36}},\
  \bibinfo {pages} {1287} (\bibinfo {year} {2012})}\BibitemShut {NoStop}%
\bibitem [{\citenamefont {CHAMEL}\ \emph {et~al.}(2013)\citenamefont {CHAMEL},
  \citenamefont {HAENSEL}, \citenamefont {ZDUNIK},\ and\ \citenamefont
  {FANTINA}}]{CHAMEL_2013}%
  \BibitemOpen
  \bibfield  {author} {\bibinfo {author} {\bibfnamefont {N.}~\bibnamefont
  {CHAMEL}}, \bibinfo {author} {\bibfnamefont {P.}~\bibnamefont {HAENSEL}},
  \bibinfo {author} {\bibfnamefont {J.~L.}\ \bibnamefont {ZDUNIK}}, \ and\
  \bibinfo {author} {\bibfnamefont {A.~F.}\ \bibnamefont {FANTINA}},\ }\href
  {\doibase 10.1142/s021830131330018x} {\bibfield  {journal} {\bibinfo
  {journal} {International Journal of Modern Physics E}\ }\textbf {\bibinfo
  {volume} {22}},\ \bibinfo {pages} {1330018} (\bibinfo {year}
  {2013})}\BibitemShut {NoStop}%
\bibitem [{\citenamefont {Kouvaris}\ and\ \citenamefont
  {Nielsen}(2015)}]{Kouvaris:2015rea}%
  \BibitemOpen
  \bibfield  {author} {\bibinfo {author} {\bibfnamefont {C.}~\bibnamefont
  {Kouvaris}}\ and\ \bibinfo {author} {\bibfnamefont {N.~G.}\ \bibnamefont
  {Nielsen}},\ }\href {\doibase 10.1103/PhysRevD.92.063526} {\bibfield
  {journal} {\bibinfo  {journal} {Phys. Rev. D}\ }\textbf {\bibinfo {volume}
  {92}},\ \bibinfo {pages} {063526} (\bibinfo {year} {2015})},\ \Eprint
  {http://arxiv.org/abs/1507.00959} {arXiv:1507.00959 [hep-ph]} \BibitemShut
  {NoStop}%
\bibitem [{\citenamefont {Mukhopadhyay}\ \emph {et~al.}(2017)\citenamefont
  {Mukhopadhyay}, \citenamefont {Atta}, \citenamefont {Imam}, \citenamefont
  {Basu},\ and\ \citenamefont {Samanta}}]{Mukhopadhyay:2016dsg}%
  \BibitemOpen
  \bibfield  {author} {\bibinfo {author} {\bibfnamefont {S.}~\bibnamefont
  {Mukhopadhyay}}, \bibinfo {author} {\bibfnamefont {D.}~\bibnamefont {Atta}},
  \bibinfo {author} {\bibfnamefont {K.}~\bibnamefont {Imam}}, \bibinfo {author}
  {\bibfnamefont {D.~N.}\ \bibnamefont {Basu}}, \ and\ \bibinfo {author}
  {\bibfnamefont {C.}~\bibnamefont {Samanta}},\ }\href {\doibase
  10.1140/epjc/s10052-017-5006-3} {\bibfield  {journal} {\bibinfo  {journal}
  {Eur. Phys. J. C}\ }\textbf {\bibinfo {volume} {77}},\ \bibinfo {pages} {440}
  (\bibinfo {year} {2017})},\ \bibinfo {note} {[Erratum: Eur.Phys.J.C 77, 553
  (2017)]},\ \Eprint {http://arxiv.org/abs/1612.07093} {arXiv:1612.07093
  [nucl-th]} \BibitemShut {NoStop}%
\bibitem [{\citenamefont {Walecka}(1974)}]{Walecka:1974qa}%
  \BibitemOpen
  \bibfield  {author} {\bibinfo {author} {\bibfnamefont {J.~D.}\ \bibnamefont
  {Walecka}},\ }\href {\doibase 10.1016/0003-4916(74)90208-5} {\bibfield
  {journal} {\bibinfo  {journal} {Annals Phys.}\ }\textbf {\bibinfo {volume}
  {83}},\ \bibinfo {pages} {491} (\bibinfo {year} {1974})}\BibitemShut
  {NoStop}%
\bibitem [{\citenamefont {Schmitt}(2015)}]{Schmitt:2014eka}%
  \BibitemOpen
  \bibfield  {author} {\bibinfo {author} {\bibfnamefont {A.}~\bibnamefont
  {Schmitt}},\ }\href {\doibase 10.1007/978-3-319-07947-9} {\emph {\bibinfo
  {title} {{Introduction to Superfluidity}: {Field-theoretical approach and
  applications}}}},\ Vol.\ \bibinfo {volume} {888}\ (\bibinfo {year} {2015})\
  \Eprint {http://arxiv.org/abs/1404.1284} {arXiv:1404.1284 [hep-ph]}
  \BibitemShut {NoStop}%
\bibitem [{\citenamefont {Li}\ \emph {et~al.}(2012{\natexlab{b}})\citenamefont
  {Li}, \citenamefont {Wang},\ and\ \citenamefont {Cheng}}]{Li:2012qf}%
  \BibitemOpen
  \bibfield  {author} {\bibinfo {author} {\bibfnamefont {X.}~\bibnamefont
  {Li}}, \bibinfo {author} {\bibfnamefont {F.}~\bibnamefont {Wang}}, \ and\
  \bibinfo {author} {\bibfnamefont {K.~S.}\ \bibnamefont {Cheng}},\ }\href
  {\doibase 10.1088/1475-7516/2012/10/031} {\bibfield  {journal} {\bibinfo
  {journal} {JCAP}\ }\textbf {\bibinfo {volume} {10}},\ \bibinfo {pages} {031}
  (\bibinfo {year} {2012}{\natexlab{b}})},\ \Eprint
  {http://arxiv.org/abs/1210.1748} {arXiv:1210.1748 [astro-ph.CO]} \BibitemShut
  {NoStop}%
\bibitem [{\citenamefont {Miao}\ \emph {et~al.}(2022)\citenamefont {Miao},
  \citenamefont {Zhu}, \citenamefont {Li},\ and\ \citenamefont
  {Huang}}]{Miao:2022rqj}%
  \BibitemOpen
  \bibfield  {author} {\bibinfo {author} {\bibfnamefont {Z.}~\bibnamefont
  {Miao}}, \bibinfo {author} {\bibfnamefont {Y.}~\bibnamefont {Zhu}}, \bibinfo
  {author} {\bibfnamefont {A.}~\bibnamefont {Li}}, \ and\ \bibinfo {author}
  {\bibfnamefont {F.}~\bibnamefont {Huang}},\ }\href@noop {} {\  (\bibinfo
  {year} {2022})},\ \Eprint {http://arxiv.org/abs/2204.05560} {arXiv:2204.05560
  [astro-ph.HE]} \BibitemShut {NoStop}%
\bibitem [{\citenamefont {Cromartie}\ \emph {et~al.}(2019)\citenamefont
  {Cromartie} \emph {et~al.}}]{Cromartie:2019kug}%
  \BibitemOpen
  \bibfield  {author} {\bibinfo {author} {\bibfnamefont {H.~T.}\ \bibnamefont
  {Cromartie}} \emph {et~al.},\ }\href {\doibase 10.1038/s41550-019-0880-2}
  {\bibfield  {journal} {\bibinfo  {journal} {Nat. Astron.}\ }\textbf {\bibinfo
  {volume} {4}},\ \bibinfo {pages} {72} (\bibinfo {year} {2019})},\ \Eprint
  {http://arxiv.org/abs/1904.06759} {arXiv:1904.06759 [astro-ph.HE]}
  \BibitemShut {NoStop}%
\bibitem [{\citenamefont {Croon}\ \emph {et~al.}(2020)\citenamefont {Croon},
  \citenamefont {McKeen},\ and\ \citenamefont {Raj}}]{Croon:2020wpr}%
  \BibitemOpen
  \bibfield  {author} {\bibinfo {author} {\bibfnamefont {D.}~\bibnamefont
  {Croon}}, \bibinfo {author} {\bibfnamefont {D.}~\bibnamefont {McKeen}}, \
  and\ \bibinfo {author} {\bibfnamefont {N.}~\bibnamefont {Raj}},\ }\href
  {\doibase 10.1103/PhysRevD.101.083013} {\bibfield  {journal} {\bibinfo
  {journal} {Phys. Rev. D}\ }\textbf {\bibinfo {volume} {101}},\ \bibinfo
  {pages} {083013} (\bibinfo {year} {2020})},\ \Eprint
  {http://arxiv.org/abs/2002.08962} {arXiv:2002.08962 [astro-ph.CO]}
  \BibitemShut {NoStop}%
\bibitem [{\citenamefont {Colpi}\ \emph {et~al.}(1986)\citenamefont {Colpi},
  \citenamefont {Shapiro},\ and\ \citenamefont
  {Wasserman}}]{PhysRevLett.57.2485}%
  \BibitemOpen
  \bibfield  {author} {\bibinfo {author} {\bibfnamefont {M.}~\bibnamefont
  {Colpi}}, \bibinfo {author} {\bibfnamefont {S.~L.}\ \bibnamefont {Shapiro}},
  \ and\ \bibinfo {author} {\bibfnamefont {I.}~\bibnamefont {Wasserman}},\
  }\href {\doibase 10.1103/PhysRevLett.57.2485} {\bibfield  {journal} {\bibinfo
   {journal} {Phys. Rev. Lett.}\ }\textbf {\bibinfo {volume} {57}},\ \bibinfo
  {pages} {2485} (\bibinfo {year} {1986})}\BibitemShut {NoStop}%
\bibitem [{\citenamefont {Sennett}\ \emph {et~al.}(2017)\citenamefont
  {Sennett}, \citenamefont {Hinderer}, \citenamefont {Steinhoff}, \citenamefont
  {Buonanno},\ and\ \citenamefont {Ossokine}}]{Sennett:2017etc}%
  \BibitemOpen
  \bibfield  {author} {\bibinfo {author} {\bibfnamefont {N.}~\bibnamefont
  {Sennett}}, \bibinfo {author} {\bibfnamefont {T.}~\bibnamefont {Hinderer}},
  \bibinfo {author} {\bibfnamefont {J.}~\bibnamefont {Steinhoff}}, \bibinfo
  {author} {\bibfnamefont {A.}~\bibnamefont {Buonanno}}, \ and\ \bibinfo
  {author} {\bibfnamefont {S.}~\bibnamefont {Ossokine}},\ }\href {\doibase
  10.1103/PhysRevD.96.024002} {\bibfield  {journal} {\bibinfo  {journal} {Phys.
  Rev. D}\ }\textbf {\bibinfo {volume} {96}},\ \bibinfo {pages} {024002}
  (\bibinfo {year} {2017})},\ \Eprint {http://arxiv.org/abs/1704.08651}
  {arXiv:1704.08651 [gr-qc]} \BibitemShut {NoStop}%
\bibitem [{\citenamefont {Abbott}\ \emph {et~al.}(2017)\citenamefont {Abbott},
  \citenamefont {Abbott}, \citenamefont {Abbott}, \citenamefont {Acernese},
  \citenamefont {Ackley}, \citenamefont {Adams}, \citenamefont {Adams},
  \citenamefont {Addesso}, \citenamefont {Adhikari}, \citenamefont {Adya},
  \citenamefont {Affeldt}, \citenamefont {Afrough}, \citenamefont {Agarwal},
  \citenamefont {Agathos}, \citenamefont {Agatsuma}, \citenamefont {Aggarwal},
  \citenamefont {Aguiar}, \citenamefont {Aiello}, \citenamefont {Ain},\ and\
  \citenamefont {Ajith}}]{PhysRevLett.119.161101}%
  \BibitemOpen
  \bibfield  {author} {\bibinfo {author} {\bibfnamefont {B.~P.}\ \bibnamefont
  {Abbott}}, \bibinfo {author} {\bibfnamefont {R.}~\bibnamefont {Abbott}},
  \bibinfo {author} {\bibfnamefont {T.~D.}\ \bibnamefont {Abbott}}, \bibinfo
  {author} {\bibfnamefont {F.}~\bibnamefont {Acernese}}, \bibinfo {author}
  {\bibfnamefont {K.}~\bibnamefont {Ackley}}, \bibinfo {author} {\bibfnamefont
  {C.}~\bibnamefont {Adams}}, \bibinfo {author} {\bibfnamefont
  {T.}~\bibnamefont {Adams}}, \bibinfo {author} {\bibfnamefont
  {P.}~\bibnamefont {Addesso}}, \bibinfo {author} {\bibfnamefont {R.~X.}\
  \bibnamefont {Adhikari}}, \bibinfo {author} {\bibfnamefont {V.~B.}\
  \bibnamefont {Adya}}, \bibinfo {author} {\bibfnamefont {C.}~\bibnamefont
  {Affeldt}}, \bibinfo {author} {\bibfnamefont {M.}~\bibnamefont {Afrough}},
  \bibinfo {author} {\bibfnamefont {B.}~\bibnamefont {Agarwal}}, \bibinfo
  {author} {\bibfnamefont {M.}~\bibnamefont {Agathos}}, \bibinfo {author}
  {\bibfnamefont {K.}~\bibnamefont {Agatsuma}}, \bibinfo {author}
  {\bibfnamefont {N.}~\bibnamefont {Aggarwal}}, \bibinfo {author}
  {\bibfnamefont {O.~D.}\ \bibnamefont {Aguiar}}, \bibinfo {author}
  {\bibfnamefont {L.}~\bibnamefont {Aiello}}, \bibinfo {author} {\bibfnamefont
  {A.}~\bibnamefont {Ain}}, \ and\ \bibinfo {author} {\bibfnamefont
  {P.}~\bibnamefont {Ajith}} (\bibinfo {collaboration} {LIGO Scientific
  Collaboration and Virgo Collaboration}),\ }\href {\doibase
  10.1103/PhysRevLett.119.161101} {\bibfield  {journal} {\bibinfo  {journal}
  {Phys. Rev. Lett.}\ }\textbf {\bibinfo {volume} {119}},\ \bibinfo {pages}
  {161101} (\bibinfo {year} {2017})}\BibitemShut {NoStop}%
\bibitem [{\citenamefont {Abbott}\ \emph {et~al.}(2020)\citenamefont {Abbott},
  \citenamefont {Abbott}, \citenamefont {Abbott}, \citenamefont {Abraham},
  \citenamefont {Acernese}, \citenamefont {Ackley}, \citenamefont {Adams},
  \citenamefont {Adhikari}, \citenamefont {Adya}, \citenamefont {Affeldt},
  \citenamefont {Agathos}, \citenamefont {Agatsuma}, \citenamefont {Aggarwal},
  \citenamefont {Aguiar}, \citenamefont {Aiello}, \citenamefont {Ain},\ and\
  \citenamefont {Ajith}}]{Abbott_2020}%
  \BibitemOpen
  \bibfield  {author} {\bibinfo {author} {\bibfnamefont {B.~P.}\ \bibnamefont
  {Abbott}}, \bibinfo {author} {\bibfnamefont {R.}~\bibnamefont {Abbott}},
  \bibinfo {author} {\bibfnamefont {T.~D.}\ \bibnamefont {Abbott}}, \bibinfo
  {author} {\bibfnamefont {S.}~\bibnamefont {Abraham}}, \bibinfo {author}
  {\bibfnamefont {F.}~\bibnamefont {Acernese}}, \bibinfo {author}
  {\bibfnamefont {K.}~\bibnamefont {Ackley}}, \bibinfo {author} {\bibfnamefont
  {C.}~\bibnamefont {Adams}}, \bibinfo {author} {\bibfnamefont {R.~X.}\
  \bibnamefont {Adhikari}}, \bibinfo {author} {\bibfnamefont {V.~B.}\
  \bibnamefont {Adya}}, \bibinfo {author} {\bibfnamefont {C.}~\bibnamefont
  {Affeldt}}, \bibinfo {author} {\bibfnamefont {M.}~\bibnamefont {Agathos}},
  \bibinfo {author} {\bibfnamefont {K.}~\bibnamefont {Agatsuma}}, \bibinfo
  {author} {\bibfnamefont {N.}~\bibnamefont {Aggarwal}}, \bibinfo {author}
  {\bibfnamefont {O.~D.}\ \bibnamefont {Aguiar}}, \bibinfo {author}
  {\bibfnamefont {L.}~\bibnamefont {Aiello}}, \bibinfo {author} {\bibfnamefont
  {A.}~\bibnamefont {Ain}}, \ and\ \bibinfo {author} {\bibfnamefont
  {P.}~\bibnamefont {Ajith}},\ }\href {\doibase 10.3847/2041-8213/ab75f5}
  {\bibfield  {journal} {\bibinfo  {journal} {The Astrophysical Journal
  Letters}\ }\textbf {\bibinfo {volume} {892}},\ \bibinfo {pages} {L3}
  (\bibinfo {year} {2020})}\BibitemShut {NoStop}%
\bibitem [{\citenamefont {Flanagan}\ and\ \citenamefont
  {Hinderer}(2008)}]{Flanagan:2007ix}%
  \BibitemOpen
  \bibfield  {author} {\bibinfo {author} {\bibfnamefont {E.~E.}\ \bibnamefont
  {Flanagan}}\ and\ \bibinfo {author} {\bibfnamefont {T.}~\bibnamefont
  {Hinderer}},\ }\href {\doibase 10.1103/PhysRevD.77.021502} {\bibfield
  {journal} {\bibinfo  {journal} {Phys. Rev. D}\ }\textbf {\bibinfo {volume}
  {77}},\ \bibinfo {pages} {021502} (\bibinfo {year} {2008})},\ \Eprint
  {http://arxiv.org/abs/0709.1915} {arXiv:0709.1915 [astro-ph]} \BibitemShut
  {NoStop}%
\bibitem [{\citenamefont {Hinderer}(2008)}]{Hinderer:2007mb}%
  \BibitemOpen
  \bibfield  {author} {\bibinfo {author} {\bibfnamefont {T.}~\bibnamefont
  {Hinderer}},\ }\href {\doibase 10.1086/533487} {\bibfield  {journal}
  {\bibinfo  {journal} {Astrophys. J.}\ }\textbf {\bibinfo {volume} {677}},\
  \bibinfo {pages} {1216} (\bibinfo {year} {2008})},\ \Eprint
  {http://arxiv.org/abs/0711.2420} {arXiv:0711.2420 [astro-ph]} \BibitemShut
  {NoStop}%
\bibitem [{\citenamefont {Perot}\ \emph {et~al.}(2020)\citenamefont {Perot},
  \citenamefont {Chamel},\ and\ \citenamefont {Sourie}}]{PhysRevC.101.015806}%
  \BibitemOpen
  \bibfield  {author} {\bibinfo {author} {\bibfnamefont {L.}~\bibnamefont
  {Perot}}, \bibinfo {author} {\bibfnamefont {N.}~\bibnamefont {Chamel}}, \
  and\ \bibinfo {author} {\bibfnamefont {A.}~\bibnamefont {Sourie}},\ }\href
  {\doibase 10.1103/PhysRevC.101.015806} {\bibfield  {journal} {\bibinfo
  {journal} {Phys. Rev. C}\ }\textbf {\bibinfo {volume} {101}},\ \bibinfo
  {pages} {015806} (\bibinfo {year} {2020})}\BibitemShut {NoStop}%
\bibitem [{\citenamefont {Most}\ \emph {et~al.}(2018)\citenamefont {Most},
  \citenamefont {Weih}, \citenamefont {Rezzolla},\ and\ \citenamefont
  {Schaffner-Bielich}}]{PhysRevLett.120.261103}%
  \BibitemOpen
  \bibfield  {author} {\bibinfo {author} {\bibfnamefont {E.~R.}\ \bibnamefont
  {Most}}, \bibinfo {author} {\bibfnamefont {L.~R.}\ \bibnamefont {Weih}},
  \bibinfo {author} {\bibfnamefont {L.}~\bibnamefont {Rezzolla}}, \ and\
  \bibinfo {author} {\bibfnamefont {J.}~\bibnamefont {Schaffner-Bielich}},\
  }\href {\doibase 10.1103/PhysRevLett.120.261103} {\bibfield  {journal}
  {\bibinfo  {journal} {Phys. Rev. Lett.}\ }\textbf {\bibinfo {volume} {120}},\
  \bibinfo {pages} {261103} (\bibinfo {year} {2018})}\BibitemShut {NoStop}%
\bibitem [{\citenamefont {Gralla}(2018)}]{Gralla:2017djj}%
  \BibitemOpen
  \bibfield  {author} {\bibinfo {author} {\bibfnamefont {S.~E.}\ \bibnamefont
  {Gralla}},\ }\href {\doibase 10.1088/1361-6382/aab186} {\bibfield  {journal}
  {\bibinfo  {journal} {Class. Quant. Grav.}\ }\textbf {\bibinfo {volume}
  {35}},\ \bibinfo {pages} {085002} (\bibinfo {year} {2018})},\ \Eprint
  {http://arxiv.org/abs/1710.11096} {arXiv:1710.11096 [gr-qc]} \BibitemShut
  {NoStop}%
\bibitem [{\citenamefont {Clesse}\ and\ \citenamefont
  {Garcia-Bellido}(2020)}]{Clesse:2020ghq}%
  \BibitemOpen
  \bibfield  {author} {\bibinfo {author} {\bibfnamefont {S.}~\bibnamefont
  {Clesse}}\ and\ \bibinfo {author} {\bibfnamefont {J.}~\bibnamefont
  {Garcia-Bellido}},\ }\href@noop {} {\  (\bibinfo {year} {2020})},\ \Eprint
  {http://arxiv.org/abs/2007.06481} {arXiv:2007.06481 [astro-ph.CO]}
  \BibitemShut {NoStop}%
\bibitem [{\citenamefont {Radice}\ \emph {et~al.}(2018)\citenamefont {Radice},
  \citenamefont {Perego}, \citenamefont {Zappa},\ and\ \citenamefont
  {Bernuzzi}}]{Radice:2017lry}%
  \BibitemOpen
  \bibfield  {author} {\bibinfo {author} {\bibfnamefont {D.}~\bibnamefont
  {Radice}}, \bibinfo {author} {\bibfnamefont {A.}~\bibnamefont {Perego}},
  \bibinfo {author} {\bibfnamefont {F.}~\bibnamefont {Zappa}}, \ and\ \bibinfo
  {author} {\bibfnamefont {S.}~\bibnamefont {Bernuzzi}},\ }\href {\doibase
  10.3847/2041-8213/aaa402} {\bibfield  {journal} {\bibinfo  {journal}
  {Astrophys. J. Lett.}\ }\textbf {\bibinfo {volume} {852}},\ \bibinfo {pages}
  {L29} (\bibinfo {year} {2018})},\ \Eprint {http://arxiv.org/abs/1711.03647}
  {arXiv:1711.03647 [astro-ph.HE]} \BibitemShut {NoStop}%
\bibitem [{\citenamefont {Riley}\ \emph {et~al.}(2019)\citenamefont {Riley},
  \citenamefont {Watts}, \citenamefont {Bogdanov}, \citenamefont {Ray},
  \citenamefont {Ludlam}, \citenamefont {Guillot}, \citenamefont {Arzoumanian},
  \citenamefont {Baker}, \citenamefont {Bilous}, \citenamefont {Chakrabarty},
  \citenamefont {Gendreau}, \citenamefont {Harding}, \citenamefont {Ho},
  \citenamefont {Lattimer}, \citenamefont {Morsink},\ and\ \citenamefont
  {Strohmayer}}]{Riley_2019}%
  \BibitemOpen
  \bibfield  {author} {\bibinfo {author} {\bibfnamefont {T.~E.}\ \bibnamefont
  {Riley}}, \bibinfo {author} {\bibfnamefont {A.~L.}\ \bibnamefont {Watts}},
  \bibinfo {author} {\bibfnamefont {S.}~\bibnamefont {Bogdanov}}, \bibinfo
  {author} {\bibfnamefont {P.~S.}\ \bibnamefont {Ray}}, \bibinfo {author}
  {\bibfnamefont {R.~M.}\ \bibnamefont {Ludlam}}, \bibinfo {author}
  {\bibfnamefont {S.}~\bibnamefont {Guillot}}, \bibinfo {author} {\bibfnamefont
  {Z.}~\bibnamefont {Arzoumanian}}, \bibinfo {author} {\bibfnamefont {C.~L.}\
  \bibnamefont {Baker}}, \bibinfo {author} {\bibfnamefont {A.~V.}\ \bibnamefont
  {Bilous}}, \bibinfo {author} {\bibfnamefont {D.}~\bibnamefont {Chakrabarty}},
  \bibinfo {author} {\bibfnamefont {K.~C.}\ \bibnamefont {Gendreau}}, \bibinfo
  {author} {\bibfnamefont {A.~K.}\ \bibnamefont {Harding}}, \bibinfo {author}
  {\bibfnamefont {W.~C.~G.}\ \bibnamefont {Ho}}, \bibinfo {author}
  {\bibfnamefont {J.~M.}\ \bibnamefont {Lattimer}}, \bibinfo {author}
  {\bibfnamefont {S.~M.}\ \bibnamefont {Morsink}}, \ and\ \bibinfo {author}
  {\bibfnamefont {T.~E.}\ \bibnamefont {Strohmayer}},\ }\href {\doibase
  10.3847/2041-8213/ab481c} {\bibfield  {journal} {\bibinfo  {journal} {The
  Astrophysical Journal Letters}\ }\textbf {\bibinfo {volume} {887}},\ \bibinfo
  {pages} {L21} (\bibinfo {year} {2019})}\BibitemShut {NoStop}%
\bibitem [{\citenamefont {{Miller}}\ \emph {et~al.}(2019)\citenamefont
  {{Miller}} \emph {et~al.}}]{2019ApJ...887L..24M}%
  \BibitemOpen
  \bibfield  {author} {\bibinfo {author} {\bibfnamefont {M.~C.}\ \bibnamefont
  {{Miller}}} \emph {et~al.},\ }\href {\doibase 10.3847/2041-8213/ab50c5}
  {\bibfield  {journal} {\bibinfo  {journal} {\apjl}\ }\textbf {\bibinfo
  {volume} {887}},\ \bibinfo {eid} {L24} (\bibinfo {year} {2019})},\ \Eprint
  {http://arxiv.org/abs/1912.05705} {arXiv:1912.05705 [astro-ph.HE]}
  \BibitemShut {NoStop}%
\bibitem [{\citenamefont {Miller}\ \emph {et~al.}(2021)\citenamefont {Miller}
  \emph {et~al.}}]{Miller_2021}%
  \BibitemOpen
  \bibfield  {author} {\bibinfo {author} {\bibfnamefont {M.~C.}\ \bibnamefont
  {Miller}} \emph {et~al.},\ }\href {\doibase 10.3847/2041-8213/ac089b}
  {\bibfield  {journal} {\bibinfo  {journal} {The Astrophysical Journal
  Letters}\ }\textbf {\bibinfo {volume} {918}},\ \bibinfo {pages} {L28}
  (\bibinfo {year} {2021})}\BibitemShut {NoStop}%
\bibitem [{\citenamefont {Croon}\ \emph {et~al.}(2022)\citenamefont {Croon},
  \citenamefont {Ipek},\ and\ \citenamefont {McKeen}}]{Croon:2022tmr}%
  \BibitemOpen
  \bibfield  {author} {\bibinfo {author} {\bibfnamefont {D.}~\bibnamefont
  {Croon}}, \bibinfo {author} {\bibfnamefont {S.}~\bibnamefont {Ipek}}, \ and\
  \bibinfo {author} {\bibfnamefont {D.}~\bibnamefont {McKeen}},\ }\href@noop {}
  {\  (\bibinfo {year} {2022})},\ \Eprint {http://arxiv.org/abs/2205.15396}
  {arXiv:2205.15396 [astro-ph.CO]} \BibitemShut {NoStop}%
\bibitem [{\citenamefont {Raffelt}(1996)}]{Raffelt:1996wa}%
  \BibitemOpen
  \bibfield  {author} {\bibinfo {author} {\bibfnamefont {G.}~\bibnamefont
  {Raffelt}},\ }\href@noop {} {\emph {\bibinfo {title} {{Stars as laboratories
  for fundamental physics}: {The astrophysics of neutrinos, axions, and other
  weakly interacting particles}}}}\ (\bibinfo {year} {1996})\BibitemShut
  {NoStop}%
\end{thebibliography}%


%
\end{document}